\documentclass[preprint,aps,prd,showpacs,preprintnumbers]{revtex4-1}

\usepackage{graphicx}
\usepackage{amsmath,amsfonts,amssymb}
\usepackage{latexsym}
\usepackage[utf8]{inputenc}

\begin{document}

\title{Polynomial $f(R)$ Palatini cosmology -- dynamical system approach}
\author{Marek Szyd{\l}owski}
\email{marek.szydlowski@uj.edu.pl}
\affiliation{Astronomical Observatory, Jagiellonian University, Orla 171, 30-244 Krakow, Poland}
\affiliation{Mark Kac Complex Systems Research Centre, Jagiellonian University, {\L}ojasiewicza 11, 30-348 Krak{\'o}w, Poland}
\author{Aleksander Stachowski}
\email{aleksander.stachowski@doctoral.uj.edu.pl}
\affiliation{Astronomical Observatory, Jagiellonian University, Orla 171, 30-244 Krakow, Poland}

\begin{abstract}
We investigate cosmological dynamics based on $f(R)$ gravity in the Palatini formulation. In this study we use the dynamical system methods. We show that the evolution of the Friedmann equation reduces to the form of the piece-wise smooth dynamical system. This system is is reduced to a 2D dynamical system of the Newtonian type. We demonstrate how the trajectories can be sewn to guarantee $C^0$ extendibility of the metric similarly as `Milne-like' FLRW spacetimes are $C^0$-extendible. We point out that importance of dynamical system of Newtonian type with non-smooth right-hand sides in the context of Palatini cosmology. In this framework we can investigate singularities which appear in the past and future of the cosmic evolution. We consider cosmological systems in both Einstein and Jordan frames. We show that at each frame the topological structures of phase space are different.
\end{abstract}

\maketitle

\section{Introduction}

Today the explanation, that the dark energy and the dark matter are some substances, is the most prominent. The opposite point of view on the description of gravity is called anti-substantialism. Extended $f(R)$ gravity models \cite{Sotiriou:2008rp,Carroll:2004de} are intrinsic or geometric models of both dark matter and dark energy. Therefore, the idea of relational gravity, in which dark matter and dark energy can be interpreted as geometric objects, is naturally realized in $f(\hat{R})$ extended gravity. The dynamical system methods in the context of investigation dynamics of $f(R)$ gravity models are used since Carroll \cite{Carroll:2004de, Borowiec:2011wd}.

The metric formulation of extended gravity model gives the fourth order field equations. This difficulty is solved by the Palatini formalism where the metric $g$ and symmetric connection $\Gamma$ are assumed to be independent dynamical variables. In this case we get a system of second order partial differential equations. The Palatini formulation is equivalent to the purely metric theory. This is consequence that the field equations for the connection $\Gamma$, firstly considered to be independent of the metric, give the Levi-Civita connection of the metric $g$.

They are many papers about the Palatini formalism. In Olmo's paper \cite{Olmo:2011uz}, the review of the Palatini $f(R)$ theories appear. In \cite{Olmo:2005hc, Olmo:2005zr} are about the scalar-tensor representation of the Palatini theories. About the existence of non-singular solutions in the Palatini gravity, they are in \cite{Barragan:2010qb, Barragan:2009sq}. In the papers: \cite{Bejarano:2017fgz, Bambi:2015zch, Olmo:2015axa, Olmo:2011np, Olmo:2011ja} are about black holes and their singularities in the Palatini approach. About the choice of a conformal frame in the Palatini gravity are in \cite{Flanagan:2003rb, Flanagan:2004bz}. Compact stars in the Starobinsky model are discussed in \cite{Pannia:2016qbj}.

The action in the Palatini approach to gravity theories $f(R)$ is postulated in the form in which the curvature scalar is treated as a function of both the metric tensor $g$ and the connection $\Gamma$, i.e. $R(g,\Gamma)=g^{\mu,\nu}R_{\mu,\nu}(\Gamma)$. Therefore, the action assumes the form
\begin{equation}
S(g,\Gamma)=\frac{1}{2}\int_\Omega \sqrt{-g}f(R)d^4 x+S_\text{matter}.
\end{equation}

After variation with respect to both dynamical variables $g$ and $\Gamma$ we obtain the Einstein field equation $(\delta g=0)$ and an additional equation which establishes some relation between the metric and the connection. If we apply the Einstein field equation this relation assumes the form of the structural equation
\begin{equation}
f'(R)R-2f(R)=T,
\end{equation}
where $T$ is trace of the energy momentum tensor.

Recently the significant and important achievements appear in the context of understanding of the Palatini theory and their application to the cosmological problem description of the evolution of the Universe \cite{Olmo:2011uz, Koivisto:2005yk, DeFelice:2010aj, Sotiriou:2008rp, Capozziello:2015wsa, Olmo:2006eh, Faraoni:2006hx}. If we considered FRW cosmological models in the Palatini framework in the Einstein frame one can obtain the exact formula for the running cosmological constant parameter \cite{Szydlowski:2017uuy}.

Cosmology is physics of the Universe but in opposite to the physical system we do not know initial conditions for the Universe. Therefore, to explain the current state of the Universe we consider all admissible physically initial conditions and study all evolutional paths for the evolution of the Universe in the universal cosmological time.

For this investigation of dynamics the tools of the dynamical system theory are especially interesting. In this approach, the evolution of the Universe is represented by trajectories in the phase space (spaces of all states of the system any time). The phase space is organized by the singular solution representing by critical points, invariant submanifolds and trajectories. Whole dynamics can be visualised in a geometrical way on the phase portrait---a phase space of all evolutional paths for all initial conditions. We are looking for attractors (repellors) in the phase space to distinguish some generic evolution scenarios for the Universe \cite{Stachowski:2016dfi, Perko:2001de}.

We describe effectively the cosmic evolution in terms of the dynamical system of the Newtonian type. In this language, the motion of a fictitious particle mimics the evolution of the universe and the potential contains all information needed for studying its dynamics. The right hand side of the system cannot be a smooth function like for the cosmological evolution governed by general relativity. However in any case they are piece-wise smooth functions. The context of application of the Palatini formalism in the investigation of cosmological dynamics discovers significance of new types of dynamical system with non-smooth right hand sides \cite{Szydlowski:2015fcq}. It is interesting that cosmological singularities can be simply characterized in terms of geometry of the potential $V(a)$, where $a$ is the scale factor \cite{Szydlowski:2015fcq}. 

In this geometrical framework singularities are manifested by a lack of analiticity of a potential itself or its derivatives with respect to the scale factor $a$ and a diagram of the potential function (or its derivatives) possesses poles at some values of scale factor $a=a_\text{sing}$. Because the potential function is a additive function of energy density components, the discontinuities appeared on a diagram of the potential $V(a)$ can be interpreted as a discontinuous jumping of a potential part. This idea that potential form posessess some part which contains jump discontinuities can be applied in different cosmological contexts. For example, it was considered to characterize singularities in phantom cosmologies \cite{Yurov:2017xjx}.

\section{Palatini formalism -- introduction}

The Palatini gravity action of $f(\hat{R})$ gravity in the Jordan frame is given by
\begin{equation}
S=S_{\text{g}}+S_{\text{m}}=\frac{1}{2}\int \sqrt{-g}f(\hat{R}) d^4 x+S_{\text{m}},\label{action}
\end{equation}
where $\hat{R}=g^{\mu\nu}\hat{R}_{\mu\nu}({\Gamma})$ is the generalized Ricci scalar and $\hat{R}_{\mu\nu}({\Gamma})$ is the Ricci tensor of a torsionless connection $\Gamma$ \cite{Allemandi:2004wn, Olmo:2011uz}. For simplifying, we assume that $8\pi G=c=1$. From action (\ref{action}), we obtain the equation of motion
\begin{equation}
f'(\hat{R})\hat{R}_{\mu\nu}-\frac{1}{2}f(\hat{R})g_{\mu\nu}=T_{\mu\nu},\label{structural}
\end{equation}
\begin{equation}
\hat{\nabla}_\alpha(\sqrt{-g}f'(\hat{R})g^{\mu\nu})=0,\label{con}
\end{equation}
where $T_{\mu\nu}=-\frac{2}{\sqrt{-g}}\frac{\delta L_{\text{m}}}{\delta g_{\mu\nu}}$ is matter energy momentum tensor and $\nabla^\mu T_{\mu\nu}=0$ and $\hat{\nabla}_\alpha$ means that the covariant derivative calculated with respect to connection $\Gamma$.

From the trace of (\ref{structural}), we get additional equation, which is called structural equation
\begin{equation}
f'(\hat{R})\hat{R}-2 f(\hat{R})=T.\label{structural2}
\end{equation}
where $T=g^{\mu\nu}T_{\mu\nu}$.

The metric $g$ is the FRW metric
\begin{equation}\label{frw}
ds^2=-dt^2+a^2(t)\left[\frac{1}{1-kr^2}dr^2+r^2(d\theta^2+\sin^2\theta d\phi^2)\right],
\end{equation}
where $a(t)$ is the scale factor, $k$ is a constant of spatial curvature ($k=0, \pm 1$), $t$ is the cosmological time.

In this paper, we assume perfect fluid with the energy-momentum tensor
\begin{equation}
T^\mu_\nu=\text{diag}(-\rho,p,p,p),
\end{equation}
where $p=w\rho$, $w=const$ is a form of the equation of state.
From the conservation equation $T_{\nu;\mu}^{\mu}=0$ we get that $\rho=\rho_0 a^{-3(1+w)}$.
In consequence trace $T$ is in the form
\begin{equation}
T=\sum_i \rho_{i,0}(3w_i-1)a(t)^{-3(1+w_i)}.
\end{equation}
We assume baryonic and dark matter $\rho_{\text{m}}$ in the form of dust $w=0$ and dark energy $\rho_\Lambda=\Lambda$ with $w=-1$.

A form of the function $f(\hat{R})$ is unknown. In this paper we assume that the polynomial form of $f(\hat{R})$ function in the form
\begin{equation}\label{lag}
f(\hat{R})=\hat{R}+\gamma \hat{R}^2.
\end{equation}
The Lagrangian (\ref{lag}) can be treated as a deviation of the $\Lambda$CDM model, by the quadratic Starobinsky term. 

A solution of the structural equation (\ref{structural2}) has the following form
\begin{equation}\label{sol}
\hat{R}=-T\equiv 4\Lambda+\rho_{\text{m},0}a^{-3}.
\end{equation}
Note that solution (\ref{sol}) has the same form in our model like in the $\Lambda$CDM model.

The Friedmann equation in our model is given by
\begin{multline}
\frac{H^2}{H_0^2}=\frac{b^2}{\left(b+\frac{d}{2}\right)^2}\left[\Omega_{\gamma}(\Omega_{\text{m},0}a^{-3}+\Omega_{\Lambda,0})^2 \frac{(K-3)(K+1)}{2b} \right. \\ 
\left. +(\Omega_\text{m,0}a^{-3}+\Omega_{\Lambda,0})
 +\frac{\Omega_{\text{r},0}a^{-4}}{b}+
\Omega_k\right],\label{friedmann2}
\end{multline}
where
$\Omega_k = -\frac{k}{H_0^2 a^2}$, 
$\Omega_{\text{r},0} = \frac{\rho_\text{r,0}}{3H_0^2}$, 
$\Omega_{\text{m},0} = \frac{\rho_\text{m,0}}{3H_0^2}$, 
$\Omega_{\Lambda,0} = \frac{\Lambda}{3H_0^2}$, 
$K = \frac{3\Omega_{\Lambda,0}}{(\Omega_\text{m,0}a^{-3}+\Omega_{\Lambda,0})}$, 
$\Omega_{\gamma} = 3\gamma H_0^2$, 
$b = f'(\hat{R})=1+2\Omega_\gamma(\Omega_\text{m,0}a^{-3}+4\Omega_{\Lambda,0})$, 
$d = \frac{1}{H}\frac{db}{dt}=-2\Omega_{\gamma}(\Omega_\text{m,0}a^{-3}+\Omega_{\Lambda,0})(3-K)$, $H_0$ is the present value of Hubble function, $\rho_{\text{r},0}$ is the present value of the energy density of radiation, $\rho_\text{m,0}$ is the present value of the density of matter. For simplicity, henceforth, we consider the model without radiation ($\rho_{\text{r},0} = 0$). Note that for $\gamma=0$, we get the $\Lambda$CDM model.

\section{Type of singularities in the Palatini model in the Jordan frame}

In cosmology many of new types of singularities were classified by Nojiri et al. \cite{Nojiri:2005sx}. This classification of the type of singularities depend on the behaviour of the scale factor $a$, the Hubble parameter $H$, the pressure $p$ and the energy density $\rho$.

\begin{itemize}
	\item Type 0: `Big crunch'. The scale factor $a$ is vanishing and $H$, $\rho$ and $p$ are blown up.
	\item Type I: `Big rip'. The scale factor $a$, $\rho$ and $p$ are blown up.
	\item Type II: `Sudden'. The scale factor $a$, $\rho$ and $H$ are finite and $\dot{H}$ and $p$ are divergent.
	\item Type III: `Big freeze'. The scale factor $a$ is finite and $H$, $\rho$ and $p$ are blown up \cite{Barrow:2004xh} or divergent \cite{BouhmadiLopez:2006fu}.
	\item Type IV. The scale factor $a$, $H$, $\rho$, $p$ and $\dot{H}$ are finite but higher derivatives of the scale factor $a$ diverge.
	\item Type V. The scale factor $a$ is finite but $\rho$ and $p$ vanish.
\end{itemize}

Following Kr{\'o}lak \cite{Krolak:1986}, type 0 and I are strong whereas type II, III and IV are weak singularities.

In our model new types of singularities appear which are not contained in the above classification. They are non-isolated singularities. It is an example of piece-wise smooth dynamical systems of cosmological origin.

Recently a physically relevant solution of general relativity of the type black hole spacetimes which admit $C^0$-metric extensions beyond the future Cauchy horizon has focused mathematicians' attention \cite{Sbierski:2015}, because this discovery is related with the fundamental issues concerning the strong cosmic censorship conjecture. In his paper Sbierski \cite{Sbierski:2015} noted that the Schwarzschild solution in the global Kruskal-Szekeres coordinates is $C^0$-extendible. 

In order Galloway and Ling \cite{Galloway:2016bej} reviewed some aspects of Sbierski’s methodology in the general relativity context of cosmological solutions, and use similar techniques to Sbierski in investigation of the $C^0$-extendibility of open FLRW cosmological models. They founded that a certain special class of open FLRW spacetimes, which we have dubbed `Milne-like', actually admits $C^0$-extension through the big bang. \cite{Galloway:2016bej, Galloway:2017qkr}. Recently Ling has showed that Milne-like spacetimes are a class of FLRW models admit $C^0$ spacetime extensions through the big bang \cite{Ling:2017uxe}.

The above mentioned fact and phase portraits suggest that models with sewn type of singularity can belong to a new class of metrics which admits $C^0$-extension like in the Milne-like model.

\begin{figure}
	\centering
	\includegraphics[width=0.7\linewidth]{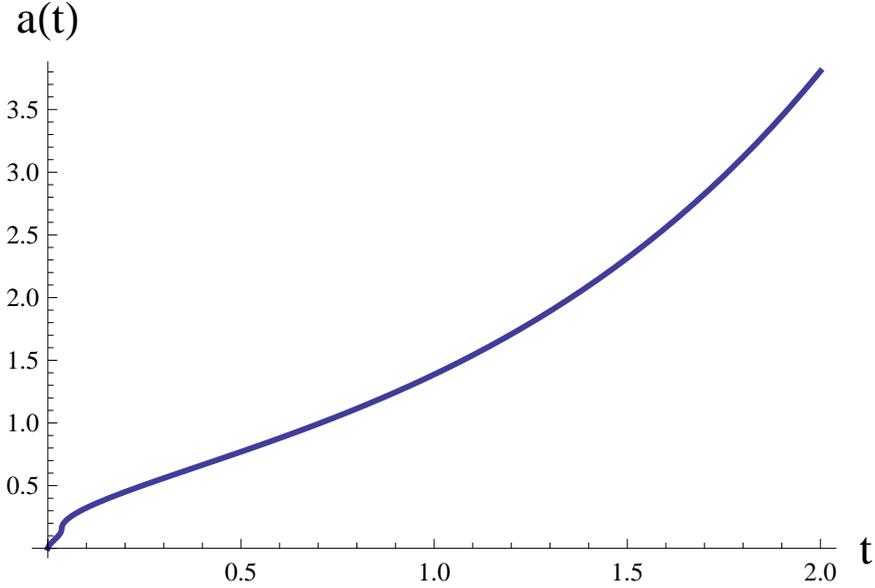}
	\caption{The illustration of the evolution of the scale factor of the model (\ref{friedmann2}) for the positive parameter $\gamma$ for the flat universe. The value of parameter $\gamma$ is chosen as $10^{-6}\frac{\text{s}^2 \text{Mpc}^2}{\text{km}^2}$. The cosmological time is expressed in $\frac{\text{s } \text{Mpc}}{100\text{ km}}$.}
	\label{fig:15}
\end{figure}

\begin{figure}
	\centering
	\includegraphics[width=0.7\linewidth]{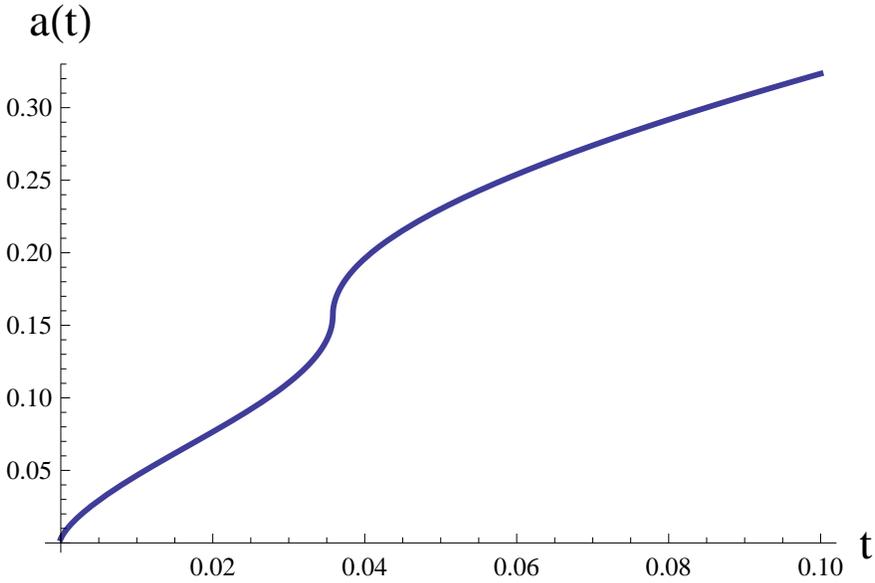}
	\caption{The illustration of the evolution of the scale factor of the model (\ref{friedmann2}) through the sewn freeze singularity (at the vertical inflection point) for the flat universe. The value of parameter $\gamma$ is chosen as $10^{-6}\frac{\text{s}^2 \text{Mpc}^2}{\text{km}^2}$. The cosmological time is expressed in $\frac{\text{s } \text{Mpc}}{100\text{ km}}$.}
	\label{fig:1}
\end{figure}

\begin{figure}
	\centering
	\includegraphics[width=0.7\linewidth]{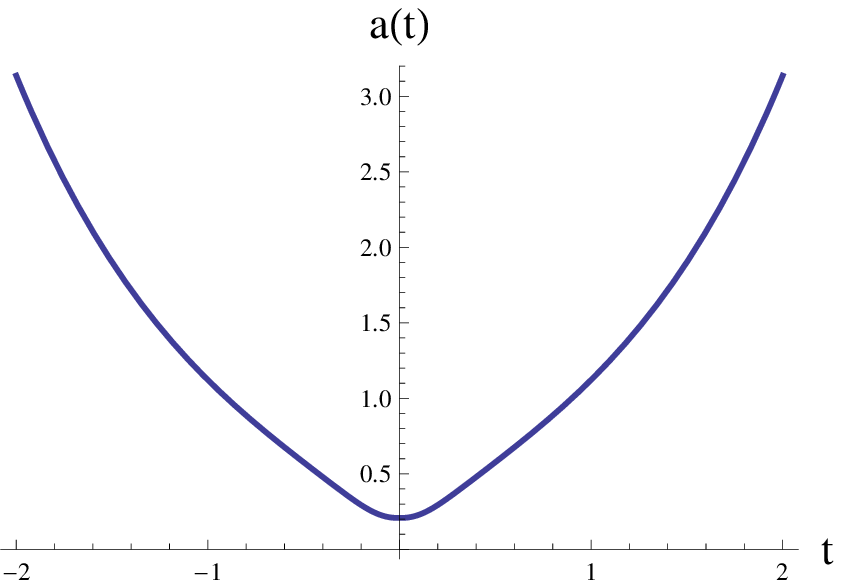}
	\caption{The illustration of the evolution of the scale factor of the model (\ref{friedmann2}) through the sewn sudden singularity (at the inflection point) for the flat universe. The model with the negative parameter $\Omega_\gamma$ has a mirror symmetry with respect to the cosmological time $t$. The bounce is at $t=0$. The value of parameter $\gamma$ is chosen as $-10^{-6}\frac{\text{s}^2 \text{Mpc}^2}{\text{km}^2}$. The cosmological time is expressed in $\frac{\text{s } \text{Mpc}}{100\text{ km}}$.}
	\label{fig:6}
\end{figure}

In our model, we find two new types of singularities, which are a consequence of the Palatini formalism: the sewn freeze and sewn sudden singularity. The freeze singularity appears when the expression $\frac{b}{b+d/2}$, in the Friedmann equation (\ref{friedmann2}), is equal the infinity. The evolution of the scale factor of the model (\ref{friedmann2}) through the sewn freeze singularity is presented in Fig.~\ref{fig:15} and \ref{fig:1}. The sewn sudden singularity appears when $\frac{b}{b+d/2}$ is equal zero. This condition is equivalent to $b=0$. The evolution of the scale factor of the model (\ref{friedmann2}) through the sewn sudden singularity is presented in Fig.~\ref{fig:6}

In our model, the sewn freeze singularity is a solution of the following algebraic equation
 \begin{equation}
 	2b+d=0
 \end{equation}
or
 \begin{equation}
-3K-\frac{K}{3\Omega_{\gamma}(\Omega_\text{m}+\Omega_{\Lambda,0})\Omega_{\Lambda,0}}+1=0,\label{k}
 \end{equation}
where $K\in [0,\text{ }3)$.

The solution of equation (\ref{k}) is
 \begin{equation}
 	 K_{\text{freeze}}=\frac{1}{3+\frac{1}{3\Omega_{\gamma}(\Omega_\text{m}+\Omega_{\Lambda,0})\Omega_{\Lambda,0}}}.\label{k2}
 \end{equation}
We obtain an expression for a value of the scale factor at the freeze singularity from equation (\ref{k2})
 \begin{equation}
 	 a_{\text{freeze}}=\left(\frac{1-\Omega_{\Lambda,0}}{8\Omega_{\Lambda,0}+\frac{1}{\Omega_{\gamma}(\Omega_\text{m}+\Omega_{\Lambda,0})}}\right)^\frac{1}{{3}}.
 \end{equation}

We get the sewn sudden singularity when $b=0$. This gets us the following algebraic equation
\begin{equation}
1+2\Omega_\gamma(\Omega_{\text{m},0}a^{-3}+4\Omega_{\Lambda,0})=0.\label{sud}
\end{equation}

From equation (\ref{sud}), we get the formula for the scale factor at a sewn sudden singularity
\begin{equation}
a_{\text{sudden}}=\left(-\frac{2\Omega_\text{m,0}}{\frac{1}{\Omega_\gamma}+8\Omega_{\Lambda,0}}\right)^{1/3}.
\end{equation}

Let the potential is
\begin{equation}\label{eq:potential}
V=-\frac{ a^2}{2}\left[\Omega_{\gamma}(\Omega_{\text{m},0}a^{-3}+4\Omega_{\Lambda,0})^2\frac{(K-3)(K+1)}{2b}+(\Omega_{\text{m},0}a^{-3}+4\Omega_{\Lambda,0})\right].
\end{equation} 
We can defined dynamical system as
\begin{align}
	a' &= y, \label{eq:ds1x}\\
	y' &= -\frac{\partial V(a)}{\partial a}, \label{eq:ds2x}
\end{align}
where $'\equiv\frac{d}{d\sigma}=\frac{b+\frac{d}{2}}{b}\frac{d}{d\tau}$ is a new parametrization of time.

We can treated dynamical system (\ref{eq:ds1x})-(\ref{eq:ds2x}) as a sewn dynamical system \cite{Hrycyna:2008gk, Ellis:2015bag}. In this case, the phase portrait is divided into two parts: the first part is for $a<a_{\text{sing}}$ and the second part is for $a>a_{\text{sing}}$. Both parts are sewn along the singularity.

For $a<a_{\text{sing}}$, we can rewritten dynamical system (\ref{eq:ds1x})-(\ref{eq:ds2x}) to the corresponding form
\begin{align}
a' &= y, \label{eq:ds3x}\\
y' &= -\frac{\partial V_1(a)}{\partial a}, \label{eq:ds4}
\end{align}
where $V_1=V(-\eta (a-a_s)+1)$ and $\eta(a)$ denotes the Heaviside function.

For $a>a_{\text{sing}}$, we get in an analogous way the following equations
\begin{align}
a' &= y, \label{eq:ds5}\\
y' &= -\frac{\partial V_2(a)}{\partial a}, \label{eq:ds6}
\end{align}
where $V_2=V \eta (a-a_s)$.

\begin{figure}
	\centering
	\includegraphics[width=0.6\linewidth]{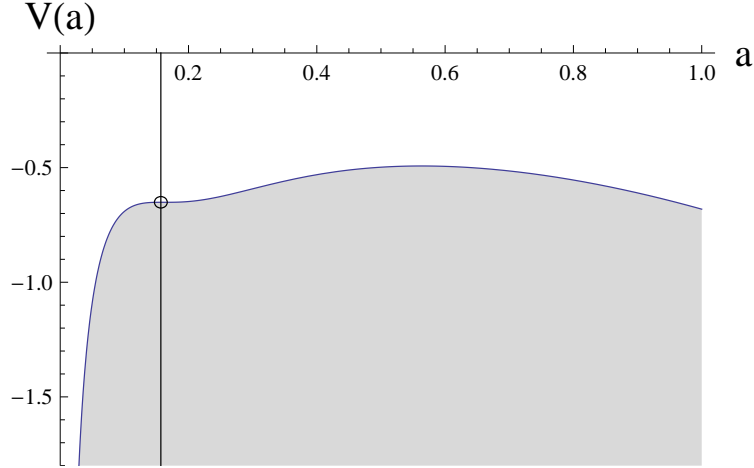}
	\caption{The diagram of the potential $V(a)$ (\ref{eq:potential}) for the positive parameter $\gamma$. The function $V(a)$ is expressed in $\frac{10^4\text{km}^2}{\text{s}^2 \text{Mpc}^2}$. The vertical line represents the sewn freeze singularity. The parameter $\gamma$ is chosen as $10^{-6}\frac{\text{s}^2 \text{Mpc}^2}{\text{km}^2}$. Note that for $a=a_\text{sing}$, $V(a)$ is undefined.}
	\label{fig:8}
\end{figure}

\begin{figure}
	\centering
	\includegraphics[width=0.6\linewidth]{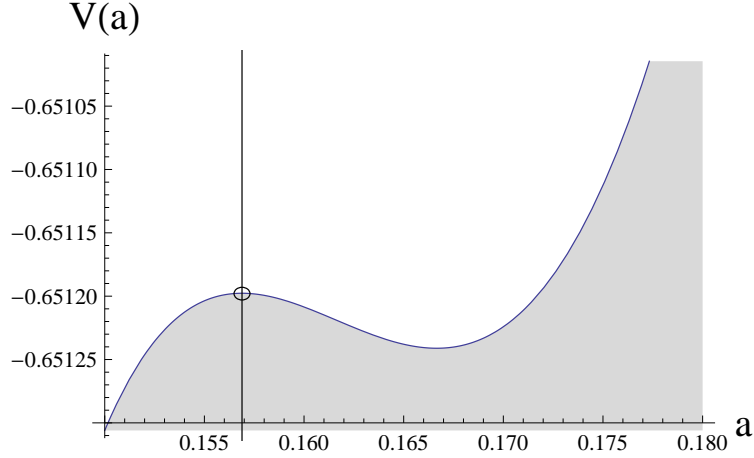}
	\caption{The diagram of potential $V(a)$ (\ref{eq:potential})for the positive parameter $\gamma$ in the neighbourhood of the freeze singularity. The function $V(a)$ is expressed in $\frac{10^4\text{km}^2}{\text{s}^2 \text{Mpc}^2}$. The vertical line represents the sewn freeze singularity. The parameter $\gamma$ is chosen as $10^{-6}\frac{\text{s}^2 \text{Mpc}^2}{\text{km}^2}$. Note that for $a=a_\text{sing}$, $V(a)$ is undefined.}
	\label{fig:12}
\end{figure}

\begin{figure}
	\centering
	\includegraphics[width=0.6\linewidth]{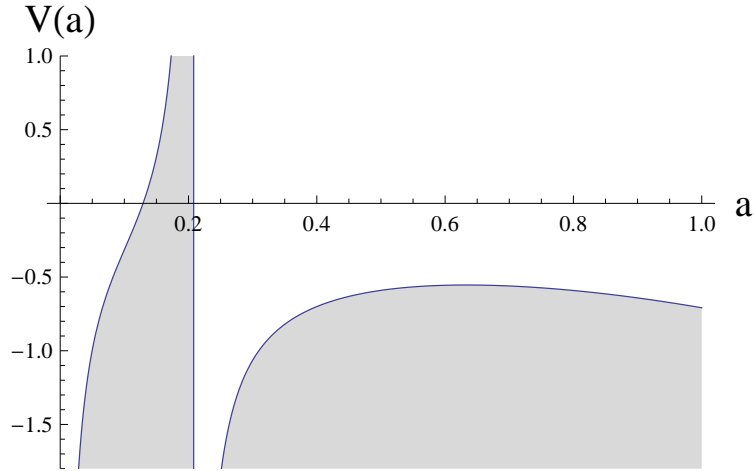}
	\caption{The diagram of the potential $V(a)$ (\ref{eq:potential}) for the negative parameter $\gamma$. The function $V(a)$ is expressed in $\frac{10^4\text{km}^2}{\text{s}^2 \text{Mpc}^2}$. The vertical line represents the sewn sudden singularity. The parameter $\gamma$ is chosen as $-10^{-6}\frac{\text{s}^2 \text{Mpc}^2}{\text{km}^2}$.}
	\label{fig:17}
\end{figure}

The diagrams of the potential function $V(a)$ (\ref{eq:potential}) are presented in Fig.~\ref{fig:8} and \ref{fig:12} for the positive parameter $\gamma$ and in Fig.~\ref{fig:17} for the negative parameter $\gamma$. The phase portraits of the system can be constructed similarly as in classical mechanics due to particle-like description of dynamics. Phase trajectories representing evolutionary paths can be obtained directly from the geometry of potential function $V(a)$ by consideration of constant energy levels $(a')^2/2 + V(a)= E=\text{const}=-k/2$. The reparametrized time parameter $\sigma$ is measured along the trajectories of the corresponding dynamical system. It has a sense of a diffeomorphic transformation beyond the singularity vertical line.

The potential function (\ref{eq:potential}) is undefined at the singularity point $a=a_{\text{sing}}$. Therefore, in phase portraits of the system in the Jordan frame there are two domains separated by a line of singularity points. These phase portraits are constructed by the application of the diffeomorphic reparametrization of cosmological time beyond this singularity line and then $C^1$ sewing of trajectories. In consequence we obtain that only one unique trajectory moves at any point of the phase space.

\begin{figure}
	\centering
	\includegraphics[width=0.6\linewidth]{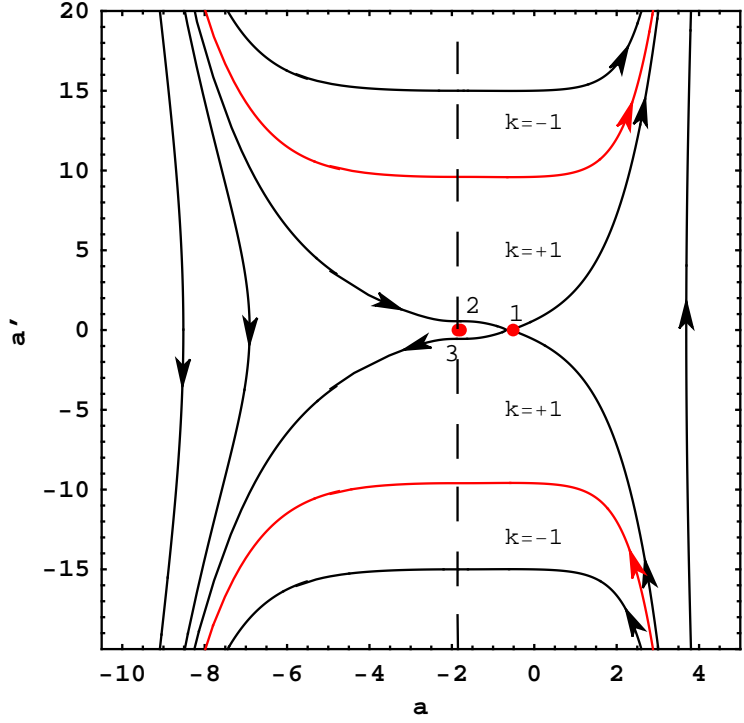}
	\caption{The phase portrait of the system (\ref{eq:ds1x}-\ref{eq:ds2x}) with the positive parameter $\Omega_{\gamma}$. The value of parameter $\gamma$ is chosen as $10^{-6}\frac{\text{s}^2 \text{Mpc}^2}{\text{km}^2}$. The value of $\Omega_{\Lambda,0}$ is chosen as 0.7 and the present value of the Hubble function is chosen as $68\frac{\text{km}}{\text{s Mpc}}$. The scale factor $a$ is presented in the natural logarithmic scale. The spatially flat universe is represented by the red trajectories. The dashed line $2b+d=0$ represents the freeze singularity. The critical points 1, 2 and 3 represent the static Einstein universes. The phase portrait belongs to the class of the sewn dynamical systems.
}
	\label{fig:2}
\end{figure}

\begin{figure}
	\centering
	\includegraphics[width=0.6\linewidth]{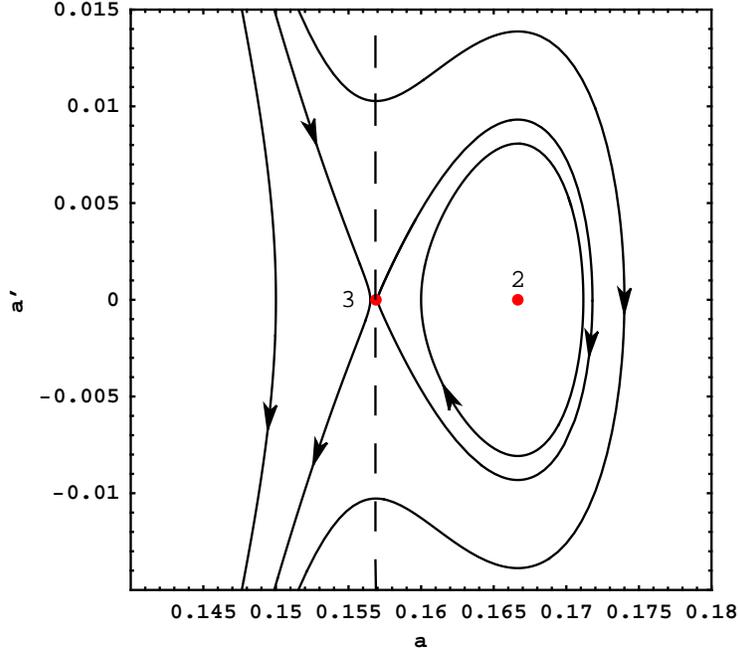}
	\caption{The zoom of the phase portrait of the system (\ref{eq:ds1x}-\ref{eq:ds2x}) with the positive parameter $\Omega_{\gamma}$ in the neighbourhood of critical points 2 and 3. Critical point 1 (see Fig.~\ref{fig:2}) is beyond the region of the diagram. The value of parameter $\gamma$ is chosen as $10^{-6}\frac{\text{s}^2 \text{Mpc}^2}{\text{km}^2}$. The value of $\Omega_{\Lambda,0}$ is chosen as 0.7 and the present value of the Hubble function is chosen as $68\frac{\text{km}}{\text{s Mpc}}$. The scale factor $a$ is presented in the natural logarithmic scale. The dashed line $2b+d=0$ represents the freeze singularity. The phase portrait belongs to the class of the sewn dynamical systems. The critical points 2 and 3 represent the static Einstein universes. Note that the existence of the homoclinic orbit which start at $t=-\infty$ and approach at $t=+\infty$. In the interior of this orbit, there are located trajectories representing oscillating cosmological models. They are free from initial and final singularities.
}
	\label{fig:11}
\end{figure}

\begin{figure}
	\centering
	\includegraphics[width=0.6\linewidth]{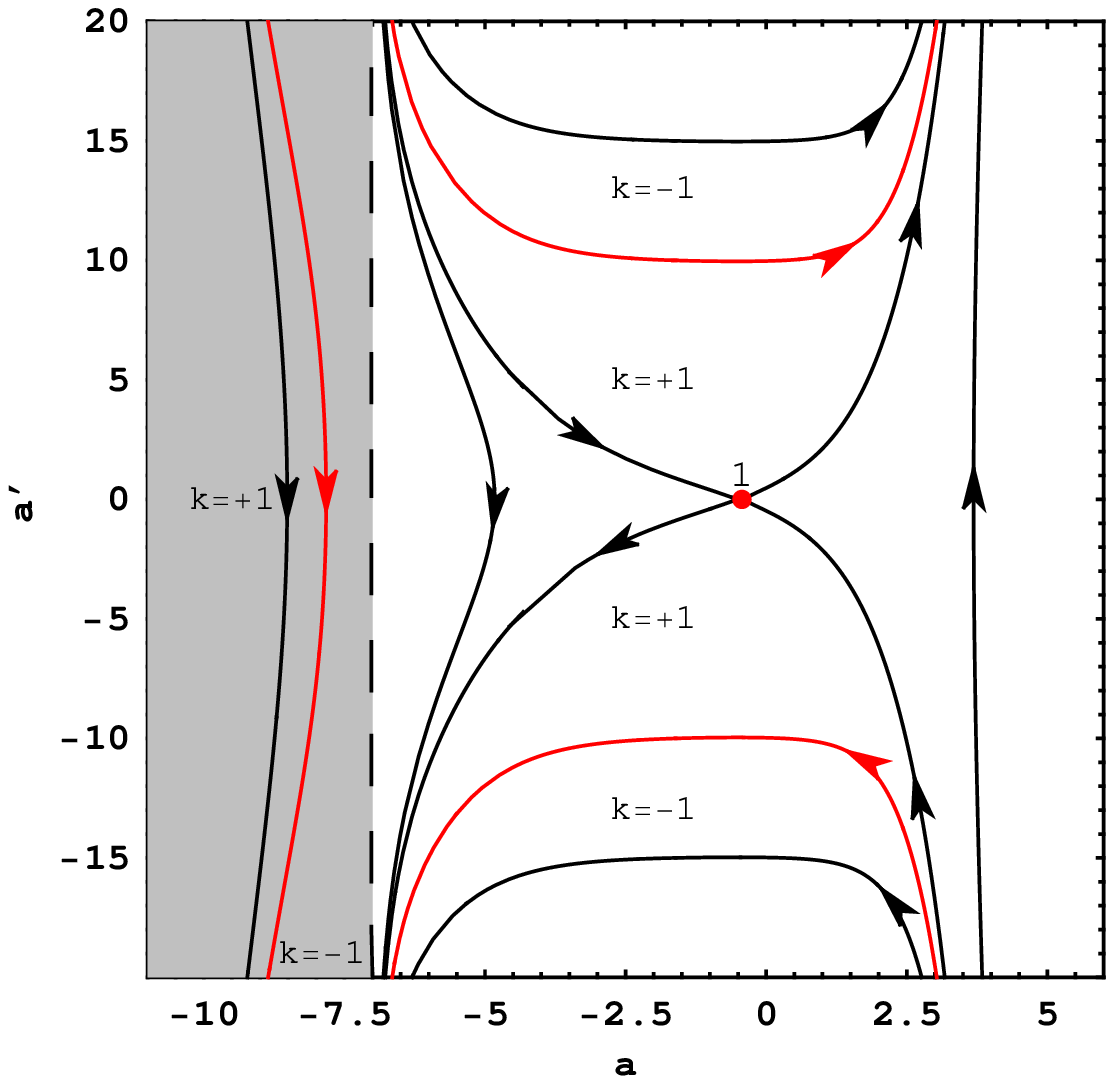}
	\caption{The phase portrait of the system (\ref{eq:ds1x}-\ref{eq:ds2x}) with the negative parameter $\Omega_{\gamma}$. The value of the parameter $\gamma$ is chosen as $-10^{-13}\frac{\text{s}^2 \text{Mpc}^2}{\text{km}^2}$. The value of $\Omega_{\Lambda,0}$ is chosen as 0.7 and the present value of the Hubble function is chosen as $68\frac{\text{km}}{\text{s Mpc}}$. The scale factor $a$ is presented in the natural logarithmic scale. The spatially flat universe is represented by the red trajectories. The dashed line separates the domain where $a<a_\text{sing}$ from the domain where $a>a_\text{sing}$. The shaded region represents trajectories with $b<0$. If we assume that $f'(R)>0$ then this region can be removed. Critical point 1 represents the static Einstein universe. The critical points at infinity, $a=a_{\text{sing}}$, $a'= \pm \infty$ are representing typical sudden singularities. The phase portrait belongs to the class of the sewn dynamical systems.
}
	\label{fig:3}
\end{figure}

The phase portraits for system (\ref{eq:ds1x}-\ref{eq:ds2x}) for positive $\Omega_{\gamma}$ are presented in Fig.~\ref{fig:2} and in Fig.~\ref{fig:11} and for negative $\Omega_{\gamma}$ is presented in Fig.~\ref{fig:3}. The line of singularity points is represented by a dashed line. 

We find that system (\ref{eq:ds1x}-\ref{eq:ds2x}) for positive $\Omega_{\gamma}$ has a sequence of three critical points located on the $a$-axis (saddle-centre-saddle sequence). To clarify the behaviour of trajectories in the neighbourhood of the saddle located at the singularity line we present the zoom of this area in Fig.~\ref{fig:11}.

In Fig.~\ref{fig:3} the critical points at infinity, $a=a_{\text{sing}}, a'= \pm \infty$ are representing typical sudden singularities. There are two types of sewn trajectories: one homoclinic orbit and infinity of periodic orbits. The homoclinic orbit starts from the neighbourhood of critical point 1, goes to the singularity at $a' = -\infty$ and after sewing with trajectory which comes from the singularity at $a' = + \infty$, finishes at the saddle point 1. The periodic orbits are situated inside the domain bounded by the homoclinic orbit. Similarly to the homoclinic orbit, the periodic orbits are sewn when going to the minus infinity singularity and going out from the plus infinity singularity. Note that these periodic orbits are possible only in the $k=+1$ universe. There are also non-periodic trajectories which lie inside the two regions bounded by the separatrices of the saddle 1. The trajectories start at $a' = -\infty$, approach saddle 1, go to the minus infinity singularity after sewing go out from the plus infinity singularity, approach saddle 1 and then continue to $a' = + \infty$. This kind of evolution is possible for the flat universe as well as $k=-1$ and $k=+1$ universes. At last in the region on the right of the separatrices of saddle 1, the trajectories start at $a' = - \infty$ and go to $a' = + \infty$ representing the evolution without sewn sudden singularity of the $k=+1$ universes.

The critical points of dynamical system (\ref{eq:ds1x}-\ref{eq:ds2x}) are completed in Table~\ref{table:1}.

\begin{table}[t]
	\caption{Critical points of dynamical system (\ref{eq:ds1x}-\ref{eq:ds2x}). They are also presented in Fig.~\ref{fig:2} and \ref{fig:11}. All three critical points represent a static Einstein universe.}
	\label{table:1}
\begin{center}
		\begin{tabular}{ccc} \hline
			no. of critical & coordinates & type \\ 
			point & of critical point & of critical point \\ \hline \hline
			1 & $\left(a=\left(\frac{8\gamma\Lambda^2-\Lambda+3H_0^2(1-8\gamma\Lambda)+\left(3H_0^2-\Lambda\right)\sqrt{(1-24\gamma\Lambda)}}{4\Lambda(1+8\gamma\Lambda)}\right)^{1/3}\text{, }a'=0\right)$ & saddle \\ \hline
			2 & $\left(a=\left(\frac{8\gamma\Lambda^2-\Lambda+3H_0^2(1-8\gamma\Lambda)-\left(3H_0^2-\Lambda\right)\sqrt{(1-24\gamma\Lambda)}}{4\Lambda(1+8\gamma\Lambda)}\right)^{1/3}\text{, }a'=0\right)$ & centre \\ \hline
			3 & $\left(a=\frac{\left(\gamma(3H_0^2-\Lambda)\right)^{1/3}}{\left(1+8\gamma\Lambda\right)^{1/3}}\text{, }a'=0\right)$ & saddle \\ \hline
			\end{tabular}
	\end{center}
\end{table}

\begin{figure}
	\centering
	\includegraphics[width=0.7\linewidth]{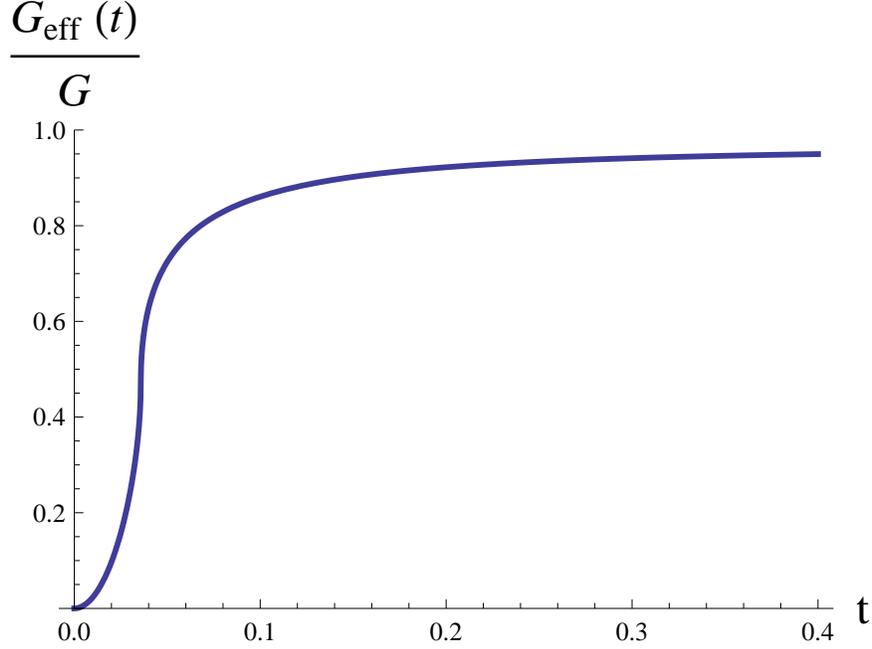}
	\caption{The evolution of $G_{\text{eff}}$ for the positive parameter $\gamma$ and the flat universe. The cosmological time $t$ is expressed in $\frac{\text{s } \text{Mpc}}{100\text{ km}}$. The parameter $\gamma$ is chosen as $10^{-6}\frac{\text{s}^2 \text{Mpc}^2}{\text{km}^2}$. Note that when $t\rightarrow\infty$ then $\frac{G_{\text{eff}}(t)}{G}\rightarrow \frac{1}{1+4\gamma \Lambda}$.}
	\label{fig:5}
\end{figure}

The action (\ref{action}) can be rewritten as
\begin{equation}
S=S_{\text{g}}+S_{\text{m}}=\frac{1}{2}\int \sqrt{-g}\phi \hat R d^4 x+S_{\text{m}},
\end{equation}
where $\phi=\frac{f(\hat R)}{\hat R}$. Let $G_{\text{eff}}$ means the effective gravitational constant. Then $\phi= \frac{1}{8\pi G_\text{eff}}$ and in the consequence $G_\text{eff}(\hat R)=\frac{\hat R}{8\pi f(\hat R)}$ and especially for $f(\hat R)=\hat R+\gamma \hat R^2$ has the following form
\begin{equation}\label{eq:geff}
\frac{G_\text{eff}(\hat R)}{G}=\frac{1}{1+\gamma \hat R}.
\end{equation}
The evolution of $G_{\text{eff}}$ is presented in Fig.~\ref{fig:5}. Note that the value of $G_{\text{eff}}$ for $t=0$ is equal zero and approaches asymptotically to the value of gravitational constant.

\section{The Palatini model in the Einstein frame}

If $f^{''}(\hat R) \neq 0 $ then action (\ref{action}) is dynamically equivalent to the first order Palatini gravitational action \cite{DeFelice:2010aj, Sotiriou:2008rp, Capozziello:2015wsa}
\begin{equation}\label{action1}
 S(g_{\mu\nu}, \Gamma^\lambda_{\rho\sigma}, \chi)=\frac{1}{2}\int\mathrm{d}^4x\sqrt{-g}\left(f^\prime(\chi)(\hat R-\chi) + f(\chi) \right) + S_m(g_{\mu\nu},\psi),
\end{equation}
Let $\Phi=f'(\chi)$ is a scalar field and $\chi=\hat R$. Then action (\ref{action1}) can be rewritten in the following form
\begin{equation}\label{actionP}
 S(g_{\mu\nu}, \Gamma^\lambda_{\rho\sigma},\Phi)=\frac{1}{2}\int\mathrm{d}^4x\sqrt{-g}\left(\Phi \hat R - U(\Phi) \right) + S_m(g_{\mu\nu},\psi),
\end{equation}
where the potential $U(\Phi)$ is defined by
\begin{equation}\label{PotentialP}
 U_f(\Phi)\equiv U(\Phi)=\chi(\Phi)\Phi-f(\chi(\Phi))%\,,\quad \bar U(\Phi)=U(\Phi)/\Phi^2.
\end{equation}
where $\Phi = \frac{d f(\chi)}{d\chi}$ and $\hat R\equiv \chi = \frac{d U(\Phi)}{d\Phi}$.

We can get from the Palatini variation of the action (\ref{actionP}) the following equations of motion
\begin{subequations}	
 \begin{align}
	\label{EOM_P}%\nonumber
	\Phi\left( \hat R_{\mu\nu} - \frac{1}{2} g_{\mu\nu} \hat R \right) &  + \frac{1}{2} g_{\mu\nu} U(\Phi) - T_{\mu\nu} = 0,\\
	\label{EOM_connectP}
	& \hat{\nabla}_\lambda(\sqrt{-g}\Phi g^{\mu\nu})=0,\\
	%\end{align}
	%
	%\begin{align}
	\label{EOM_scalar_field_P}
	  \hat R & - U^\prime(\Phi) =0.%\\
	\end{align}
\end{subequations}
From equation (\ref{EOM_connectP}), we get that the connection $\hat \Gamma$ is a metric connection for a new metric $\bar g_{\mu\nu}=\Phi g_{\mu\nu}$; thus $\hat R_{\mu\nu}=\bar R_{\mu\nu}, \bar R= \bar g^{\mu\nu}\bar R_{\mu\nu}=\Phi^{-1} \hat R$ and $\bar g_{\mu\nu}\bar R=\ g_{\mu\nu}\hat R$.
The $g$-trace of (\ref{EOM_P}) gives a new structural equation
\begin{equation}\label{struc2}
  2U(\Phi)-U'(\Phi)\Phi=T.
\end{equation}

Now equations (\ref{EOM_P}) and (\ref{EOM_scalar_field_P}) get the following form
	\begin{align}
	\label{EOM_P1}
	 \bar R_{\mu\nu} - \frac{1}{2} \bar g_{\mu\nu} \bar R  &  = \bar T_{\mu\nu}-\frac{1}{2} \bar g_{\mu\nu} \bar U(\Phi),\\
	%\end{align}
 	%
	%\begin{align}
	\label{EOM_scalar_field_P1}
	  \Phi\bar R & - (\Phi^2\,\bar U(\Phi))^\prime =0,
	\end{align}
where $\bar U(\phi)=U(\phi)/\Phi^2$, $\bar T_{\mu\nu}=\Phi^{-1}T_{\mu\nu}$ and the structural equation can be replaced by
\begin{equation}\label{EOM_P1c}
 \Phi\,\bar U^\prime(\Phi) + \bar T = 0\,.
\end{equation}
In consequence, the action for the metric $\bar g_{\mu\nu }$ and scalar field $\Phi$ is given in the following form
\begin{equation}\label{action2}
 S(\bar g_{\mu\nu},\Phi)=\frac{1}{2}\int\mathrm{d}^4x\sqrt{-\bar g}\left(\bar R- \bar U(\Phi) \right) + S_m(\Phi^{-1}\bar g_{\mu\nu},\psi),
\end{equation}
where a non-minimal coupling is between $\Phi$ and $\bar g_{\mu\nu}$
\begin{equation}\label{em_2}
    \bar T^{\mu\nu} =
-\frac{2}{\sqrt{-\bar g}} \frac{\delta}{\delta \bar g_{\mu\nu}}S_m = (\bar\rho+\bar p)\bar u^{\mu}\bar u^{\nu}+ \bar p\bar g^{\mu\nu}=\Phi^{-3}T^{\mu\nu}~,
%\left(\sqrt{-g} {\cal{L}}_{\rm m} \right)
\end{equation}
$\bar u^\mu=\Phi^{-\frac{1}{2}}u^\mu$, $\bar\rho=\Phi^{-2}\rho,\ \bar p=\Phi^{-2}p$, $\bar T_{\mu\nu}= \Phi^{-1}T_{\mu\nu}, \ \bar T= \Phi^{-2} T$ \cite{Capozziello:2015wsa, Dabrowski:2008kx}.

The FRW metric case, metric $\bar g_{\mu\nu}$ has the following form
\begin{equation}\label{frwb}
d\bar s^2=-d\bar t^2+\bar a^2(t)\left[dr^2+r^2(d\theta^2+\sin^2\theta d\phi^2)\right],
\end{equation}
where $d\bar t=\Phi(t)^{\frac{1}{2}}dt$ and new scale factor $\bar a(\bar t)=\Phi(\bar t)^{\frac{1}{2}}a(\bar t) $. 	
Because we assume the barotropic matter, the cosmological equations are given by
\begin{equation}\label{frwb2}
3\bar H^2= \bar \rho_\Phi + \bar\rho _\text{m}, \qquad 6\frac{\ddot{\bar a}}{\bar a}=2\bar\rho_\Phi -\bar{\rho}_\text{m} (1+3w)
\end{equation}
where
\begin{equation}\label{frwb3}
\bar\rho_\Phi=\frac{1}{2}\bar U(\Phi),\qquad \bar{\rho}_{\text{m}}=\rho_0\bar a^{-3(1+w)}\Phi^{\frac{1}{2}(3w-1)}\end{equation}
and
$w=\bar p_{\text{m}} / \bar\rho_{\text{m}}= p_{\text{m}} / \rho_{\text{m}}$. The conservation equation gets the following form
\begin{equation}\label{frwb4}
\dot{\bar{\rho}}_{\text{m}}+3\bar H\bar{\rho}_{\text{m}}(1+w)=-\dot{\bar{\rho}}_\Phi.
\end{equation}

In the case of the Starobinsky--Palatini model the potential $\bar U$ is described by the following formula
\begin{equation}
\bar U(\Phi)=2\bar\rho_\Phi(\Phi)=\left(\frac{1}{4\gamma}+2\lambda\right)\frac{1}{\Phi^2}-\frac{1}{2\gamma}\frac{1}{\Phi}+
\frac{1}{4\gamma}.
\end{equation}

\begin{figure}
\centering
\includegraphics[width=0.6\textwidth]{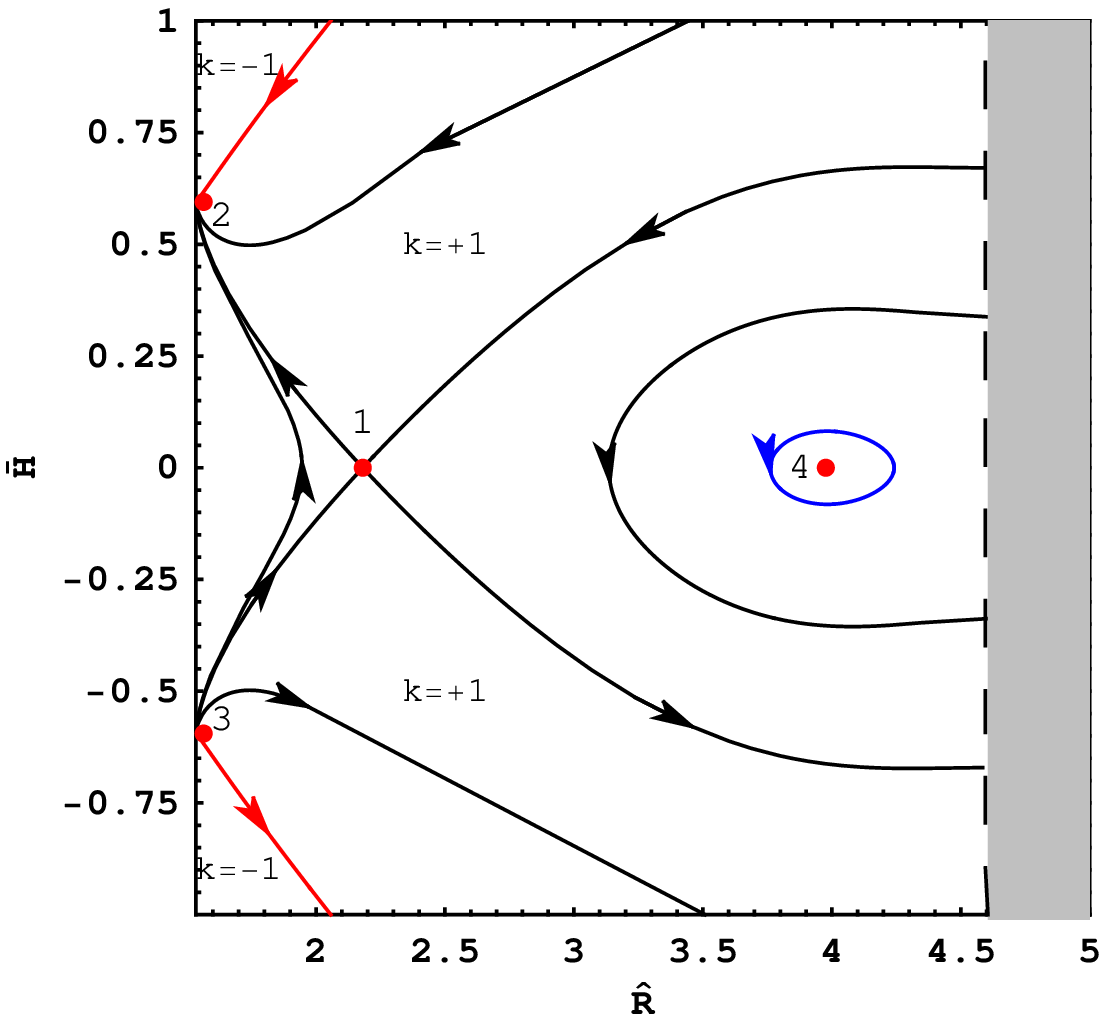}
\caption{The phase portrait of system (\ref{dynsys3})-(\ref{dynsys4}). There are four critical points: point 1 represents the Einstein universe, point 2 represents the stable de Sitter universe, point 3 represents the unstable de Sitter universe and point 4 represents the Einstein universe. The value of the parameter $\gamma$ is chosen as $10^{-6}\frac{\text{s}^2 \text{Mpc}^2}{\text{km}^2}$. The value of $\Omega_{\Lambda,0}$ is chosen as 0.7 and the present value of the Hubble function is chosen as $68\frac{\text{km}}{\text{s Mpc}}$. The values of the Hubble function are given in $\frac{100\text{km}}{\text{s Mpc}}$ and the values of the Ricci scalar are given in $\frac{10^4\text{km}^2}{\text{s}^2 \text{Mpc}^2}$ in the natural logarithmic scale. The gray colour represents the non-physical domain. The dashed line represents the generalized sudden singularity. Note that for the Starobinsky--Palatini model in the Einstein frame for the positive parameter $\gamma$, the sewn freeze singularity is replaced by the generalized sudden singularity. A typical trajectory in the neighbourhood of trajectory of the flat model (represented by the red trajectory) starts from the generalized sudden singularity then then goes to the de Sitter attractor. The position of this attractor is determined by the cosmological constant parameter. Oscillating models (blue trajectory) are situated around critical point 4.
}
\label{phase:2}
\end{figure}

\begin{figure}
\centering
\includegraphics[width=0.6\textwidth]{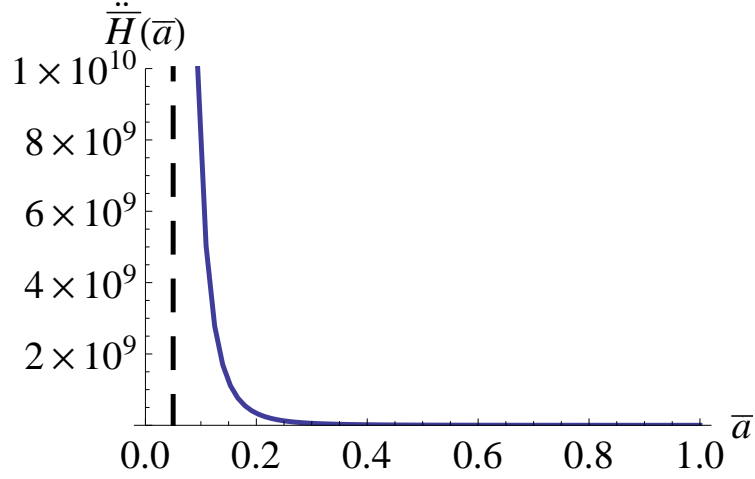}
\caption{The relation $\ddot{H}(\bar a)$ for the Palatini formalism in the Einstein frame. The value of the parameter $\gamma$ is chosen as $10^{-9}\frac{\text{s}^2 \text{Mpc}^2}{\text{km}^2}$. The values of the $\ddot{H}(\bar a)$ are given in $\frac{\text{km}^3}{\text{s}^3 \text{Mpc}^3}$. The dashed line represents the generalized sudden singularity. Note that, in the generalized sudden singularity, $H$ and $\dot H$ are finite but $\ddot H$ and its derivatives are diverge.
}
\label{fig:13}
\end{figure}

\begin{figure}
	\centering
	\includegraphics[width=0.7\linewidth]{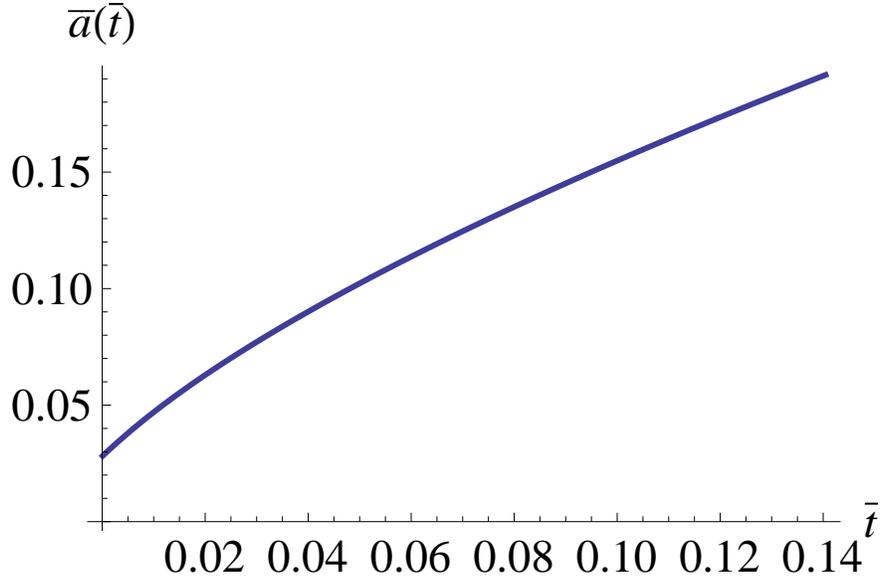}
	\caption{The illustration of the evolution of the scale factor for the Palatini formalism in the Einstein frame for the flat universe. The value of parameter $\gamma$ is chosen as $10^{-9}\frac{\text{s}^2 \text{Mpc}^2}{\text{km}^2}$. The cosmological time is expressed in $\frac{\text{s } \text{Mpc}}{\text{km}}$. Note that the evolution of the scale factor begins from the finite value different from zero.}
	\label{fig:14}
\end{figure}

\begin{figure}
\centering
\includegraphics[width=0.6\textwidth]{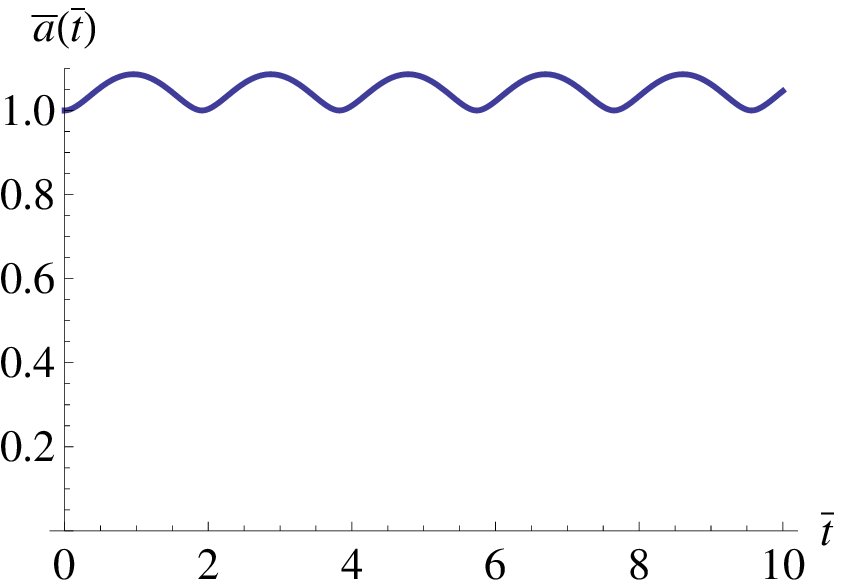}
\caption{The diagram presents the evolution of the scale factor for trajectory of the oscillating orbit in the neighbourhood of critical point 4 (see Fig.~\ref{phase:2}). The cosmological time is expressed in $\frac{\text{s } \text{Mpc}}{100\text{ km}}$. Here, $a_\text{min}=1$.
}
\label{fig:9}
\end{figure}

The cosmological equation for the Starobinsky--Palatini model in the Einstein frame can be rewritten to the form of the dynamical system with the Hubble parameter $\bar H(\bar t)$ and the Ricci scalar $\hat R(\bar t)$ as variables
\begin{multline}
\dot {\bar H}(\bar t)=\frac{1}{6\text{  }(1+2 \gamma \hat R(\bar t))^2}\\ \left( 6 \Lambda -6 \bar H(\bar t)^2 (1+2 \gamma \hat R(\bar t))^2+ \hat R(\bar t) (-1+24 \gamma  \Lambda +\gamma (1+24 \gamma
 \Lambda ) \hat R(\bar t))\right),\label{dynsys3}
 \end{multline}
 \begin{equation}
\dot {\hat R}(\bar t)=-\frac{3}{(-1+\gamma \hat R(\bar t)) }\text{  }\bar H(\bar t)(1+2 \gamma \hat R(\bar t))
 \left(4 \Lambda +\hat R(\bar t) \left(-1+16 \gamma \Lambda +16 \gamma ^2 \Lambda
 \hat R(\bar t)\right)\right),\label{dynsys4}
 \end{equation}
 where a dot denotes the differentiation with respect to the time $\bar t$. The phase portrait for dynamical system (\ref{dynsys3})-(\ref{dynsys4}) is presented in Fig.~\ref{phase:2}. Here, the periodic orbits appear around critical point 4. In the Starobinsky--Palatini model in the Einstein frame appears the generalized sudden singularity, for which $H$ and $\dot H$ are finite but $\ddot H$ and its derivatives are diverge (see Fig.~\ref{fig:13}). The evolution of the scale factor begins from the finite value different from zero (see Fig.~\ref{fig:14}). In terms of the scale factor, at the singularity for the finite value of the scale factor $\bar{a}$, a third time derivative (and higher orders) of the scale factor in Einstein frame blow up, while first and second order time derivatives behaves regularly. The evolution of the scale factor for one of these periodic orbits is presented in Fig.~\ref{fig:9}. When matter is negligible then the inflation appears. In this case $a\approx a_0\exp\left(\frac{t}{4}\sqrt{\frac{1+\sqrt{1-32\gamma\Lambda}}{\gamma}}\right)$, where $a_0=a(0)$ and $R(t)\approx\frac{1-16\gamma\Lambda+\sqrt{1-32\gamma\Lambda}}{32\gamma^2\Lambda}$ \cite{Szydlowski:2017tbc}. If $\gamma>\frac{1}{36\Lambda}$ then the non-physical domain appears for $\hat R<\frac{1-16\gamma\Lambda+\sqrt{1-32\gamma\Lambda}}{32\gamma^2\Lambda}$ for which $\rho_\text{m}<0$.

\begin{figure}
	\centering
	\includegraphics[width=0.7\linewidth]{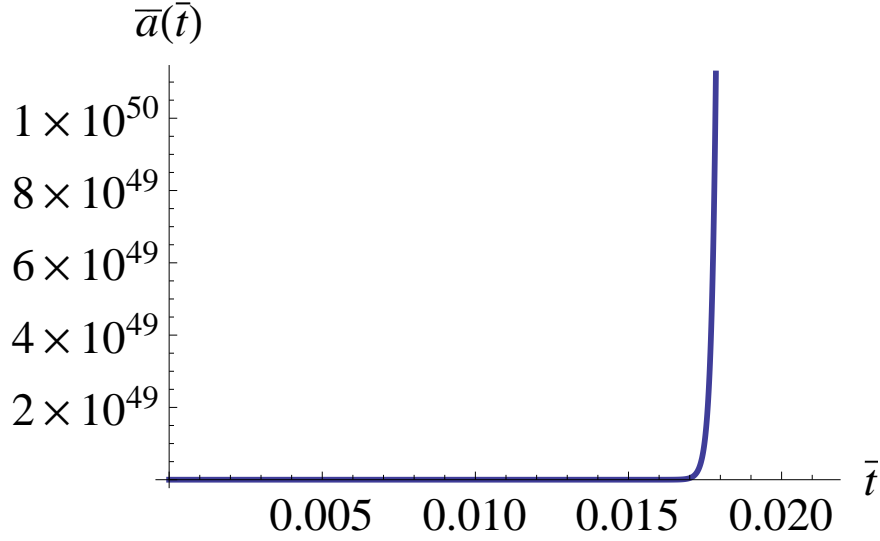}
	\caption{The illustration of the evolution of the scale factor for the Palatini formalism in the Einstein frame for the flat universe, when matter is negligible. The value of parameter $\gamma$ is chosen as $10^{-9}\frac{\text{s}^2 \text{Mpc}^2}{\text{km}^2}$. The cosmological time is expressed in $\frac{\text{s } \text{Mpc}}{\text{km}}$. Note that when matter is negligible then the inflation appears. In this case number of e-folds is equal 50.}
	\label{fig:19}
\end{figure}

\begin{figure}
\centering
\includegraphics[width=0.6\textwidth]{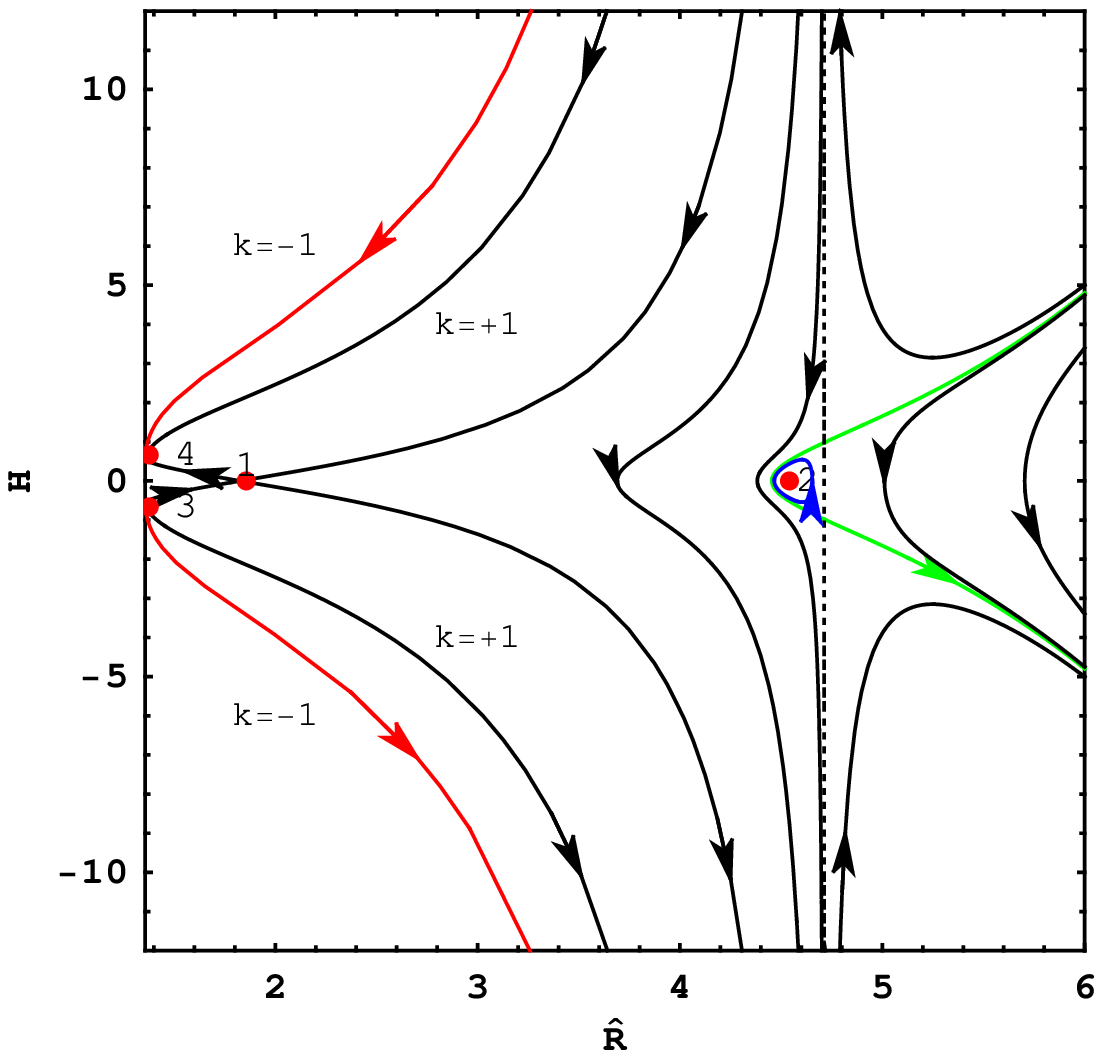}
\caption{The phase portrait of system (\ref{dynsys1})-(\ref{dynsys2}). There are four critical points: point 1 and 2 represent the Einstein universe, point 3 represents the unstable de Sitter universe, point 4 represents the stable de Sitter universe. For illustration the value of the parameter $\gamma$ is chosen as $10^{-6}\frac{\text{s}^2 \text{Mpc}^2}{\text{km}^2}$. The value of $\Omega_{\Lambda,0}$ is chosen as 0.7 and the present value of the Hubble function is chosen as $68\frac{\text{km}}{\text{s Mpc}}$. The values of the Hubble function are given in $\frac{100\text{km}}{\text{s Mpc}}$ and the values of the Ricci scalar are given in $\frac{10^4\text{km}^2}{\text{s}^2 \text{Mpc}^2}$ in the natural logarithmic scale. The dotted line, representing a line of discontinuity, separates the domain where $\hat R<\hat R_\text{sing}=\hat R(a_\text{sing})$ from the domain where $\hat R>\hat R_\text{sing}=\hat R(a_\text{sing})$. Note that oscillating models exist (blue trajectory) and are situated around critical point 2. They are representing oscillating models without the initial and final singularities. The green line represents the separatrix trajectory, which represents the only case for which the trajectory can pass from the left side of the phase portrait to the right one without appearing of the sewn freeze singularity during the evolution. It joins saddle points in a circle at the infinity. This line separates trajectories going to the freeze singularity from the bouncing solutions. For this case $\Omega_\text{k}=-\Omega_{\gamma}(\Omega_{\text{m},0}a^{-3}+4\Omega_{\Lambda,0})^2\frac{(K-3)(K+1)}{2b}-(\Omega_{\text{m},0}a^{-3}+4\Omega_{\Lambda,0})$ when $a=a_\text{sing}$.
}
\label{phase:1}
\end{figure}

\begin{figure}
\centering
\includegraphics[width=0.6\textwidth]{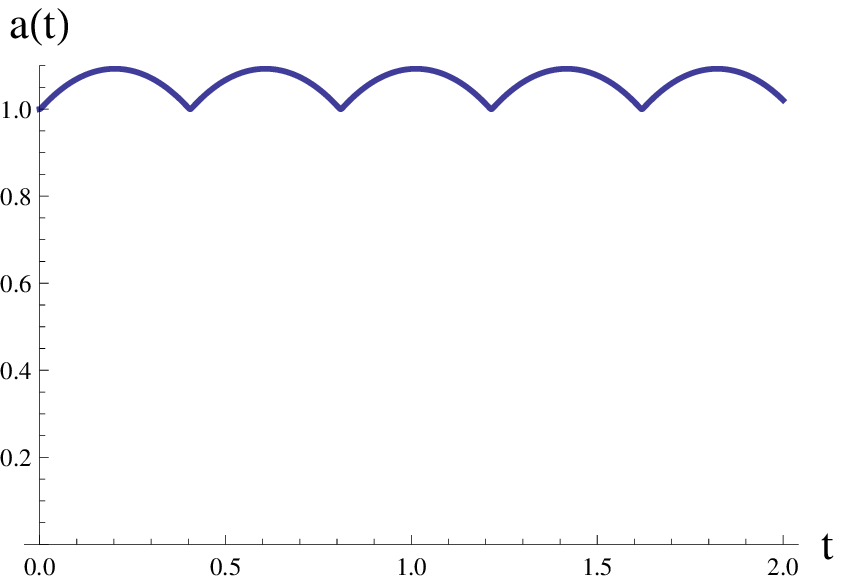}
\caption{The diagram presents the evolution of the scale factor for trajectory of the oscillating orbit in the neighbourhood of critical point 2 (see Fig.~\ref{phase:1}). The cosmological time is expressed in $\frac{\text{s } \text{Mpc}}{100\text{ km}}$. Here, $a_\text{min}=1$.
}
\label{fig:7}
\end{figure}

For comparison of the dynamical system in the both frames, we obtain dynamical system for the Starobinsky--Palatini model in the Jordan frame in the variables $H(t)$ and $\hat R(t)$ 
\begin{multline}
\dot{H}(t)=-\frac{1}{6} \left[ 6\left(2 \Lambda +H(t)^2\right)+\hat R(t)+\frac{18 (1+8 \gamma  \Lambda ) \left(\Lambda -H(t)^2\right)}{-1-12
\gamma  \Lambda +\gamma  \hat R(t)} \right. \\
\left. -\frac{18(1+8 \gamma \Lambda ) H(t)^2}{1+2 \gamma \hat R(t)} \right],\label{dynsys1}
\end{multline}
\begin{equation}
\dot {\hat R}(t)=-3 H(t)(\hat R(t)-4 \Lambda ),\label{dynsys2}
\end{equation}
where a dot means the differentiation with respect to time $t$.
The phase portrait for dynamical system (\ref{dynsys1})-(\ref{dynsys2}) is shown in Fig.~\ref{phase:1}. This phase portrait represent all evolutionary paths of the system in the Jordan frame without adopting the time reparametrization. Along the trajectories is measured original cosmological time $t$. The oscillating orbits appear around critical point 4 (see Fig.~\ref{phase:1}). The evolution of the scale factor for one of these periodic orbits is presented in Fig.~\ref{fig:7}.

\begin{figure}
\centering
\includegraphics[width=0.6\textwidth]{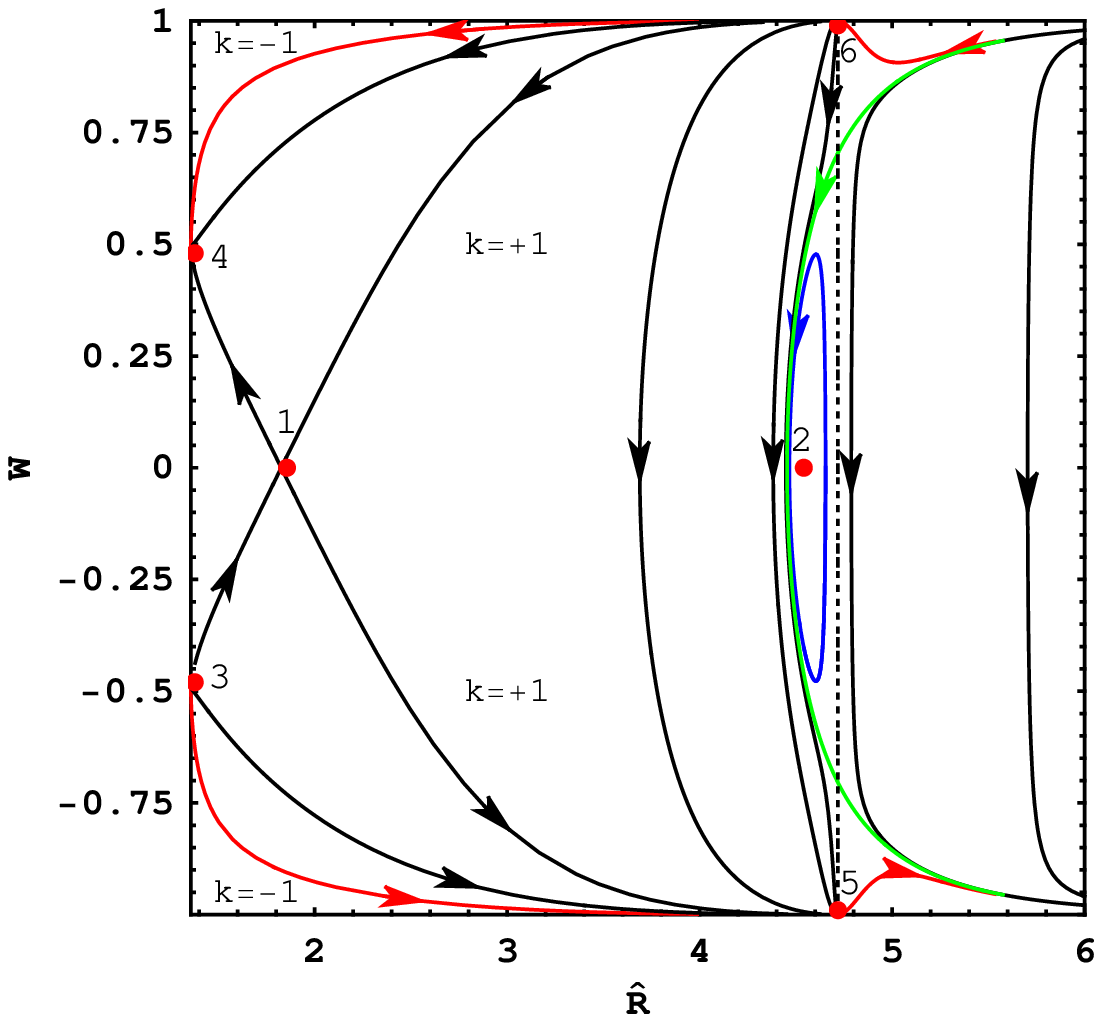}
\caption{The phase portrait of system (\ref{dynsys1x})-(\ref{dynsys2x}). There are four critical points: point 1 and 2 represent the Einstein universe, point 3 represents the unstable de Sitter universe, point 4 represents the stable de Sitter universe. For illustration the value of the parameter $\gamma$ is chosen as $10^{-6}\frac{\text{s}^2 \text{Mpc}^2}{\text{km}^2}$. The value of $\Omega_{\Lambda,0}$ is chosen as 0.7 and the present value of the Hubble function is chosen as $68\frac{\text{km}}{\text{s Mpc}}$. The Ricci scalar are given in $\frac{10^4\text{km}^2}{\text{s}^2 \text{Mpc}^2}$ in the natural logarithmic scale. The dotted line, representing a line of discontinuity, separates the domain where $\hat R<\hat R_\text{sing}=\hat R(a_\text{sing})$ from the domain where $\hat R>\hat R_\text{sing}=\hat R(a_\text{sing})$. Point 5 and 6 represent points where the right and left side of the phase space is sewn (some trajectories pass through the sewn singularity--points 5 and 6). Oscillating models (blue trajectory) are situated around critical point 2. The green line represents the special trajectory, which represents the only case for which the trajectory can pass from the left side of the phase portrait to the right one without appearing of the sewn freeze singularity during the evolution. It joins saddle points in a circle at the infinity. This line separates trajectories going to the freeze singularity from the bouncing solutions. For this case $\Omega_\text{k}=-\Omega_{\gamma}(\Omega_{\text{m},0}a^{-3}+4\Omega_{\Lambda,0})^2\frac{(K-3)(K+1)}{2b}-(\Omega_{\text{m},0}a^{-3}+4\Omega_{\Lambda,0})$ when $a=a_\text{sing}$. The dotted line 
}
\label{phase:3}
\end{figure}

For a deeper analysis of the behaviour of the trajectories of system (\ref{dynsys1})--(\ref{dynsys2}) in the infinity, we introduce variables $\hat R$ and $W=\frac{H}{\sqrt{1+H}}$ and rewrite equations (\ref{dynsys1})--(\ref{dynsys2}) in these variables. Then we get the following dynamical system
\begin{align}
\dot W(t) &=\frac{\dot{H}(t)}{\left(1+H(t)^2\right)^{3/2}}=-\frac{1}{6} \left[ 6\left(2 \Lambda +\frac{W(t)^2}{1-W(t)^2}\right)+\hat R(t)+\frac{18 (1+8 \gamma \Lambda ) \left(\Lambda -\frac{W(t)^2}{1-W(t)^2}\right)}{-1-12
\gamma \Lambda +\gamma \hat R(t)} \right. \nonumber \\
& \quad \left. -\frac{18(1+8 \gamma \Lambda ) \frac{W(t)^2}{1-W(t)^2}}{1+2 \gamma \hat R(t)} \right],\label{dynsys1x} \\
\dot {\hat R}(t) &=-3 \frac{W(t)}{\sqrt{1-W(t)^2}}(\hat R(t)-4 \Lambda ).\label{dynsys2x}
\end{align}
The phase portrait for dynamical system (\ref{dynsys1x})-(\ref{dynsys2x}) is presented in Fig.~\ref{phase:3}. This portrait is a good illustration how trajectories are sewn at the points at infinity (points 5 and 6).
If we consider expanding models situated on the upper part of the domain, $W$ is positive, all the trajectories passing through point 6. This continuation of trajectories is the class of $C^0$. The singularity line is representing the freeze type of singularity. There is some differences in the behaviour of trajectories of the same model represented in Fig.~\ref{fig:2} and Figs. \ref{phase:1} and \ref{phase:3}. While the continuation on the singularity line in Fig.~\ref{fig:2} is smooth of $C^1$ class and the Cachy problem is correctly solved in Fig.~\ref{phase:1} and \ref{phase:3} all trajectories from separated regions focused at the degenerated point 6 (and point 5 for contracting models) represent the freeze type of singularity. It has a consequence for solution of the Cauchy problem. Therefore the representation of dynamics in the reparametrized time seems to be more suitable than in the original cosmological time.

\begin{figure}
\centering
\includegraphics[width=0.6\textwidth]{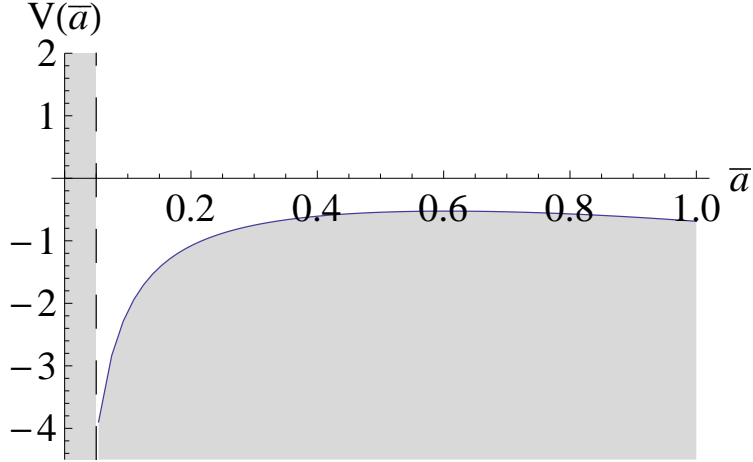}
\caption{The potential $V(\bar a)$ for the Palatini formalism in the Einstein frame. The value of the parameter $\gamma$ is chosen as $10^{-9}\frac{\text{s}^2 \text{Mpc}^2}{\text{km}^2}$. The values of the $V(\bar a)$ are given in $\frac{10^4\text{km}^2}{\text{s}^2 \text{Mpc}^2}$. The dashed line represents the generalized sudden singularity. The value of the potential at the singularity is finite.
}
\label{fig:10}
\end{figure}

For the equations (\ref{dynsys3})--(\ref{dynsys4}) and (\ref{dynsys1})--(\ref{dynsys2}), we can find the first integrals. In the case of equations (\ref{dynsys3})--(\ref{dynsys4}), the first integral has the following form
\begin{equation}
\bar H(\bar t)^2+ \Lambda -\frac{\hat R(\bar t) (2+\gamma \hat R(\bar t))}{6 (1+2 \gamma \hat R(\bar t))^2}+
\frac{k}{2\bar a^2}=0.\label{firstintegral1a}
\end{equation}
Because 
\begin{equation}
\bar a=\sqrt{\frac{C_0 (1+2 \gamma \hat R(\bar t))}{2e^{-\frac{\arctan \left(\frac{-1+16 \gamma
 \Lambda + 32 \gamma ^2 \Lambda \hat R(\bar t)}{\sqrt{-1+32 \gamma \Lambda }}\right)}{3 \sqrt{-1+32 \gamma \Lambda }}} \sqrt{4 \Lambda +\hat R(\bar t) \left(-1+16
\gamma \Lambda +16 \gamma ^2 \Lambda \hat R(\bar t)\right)}}},
\end{equation}
where $C_0=\frac{ \bar a_0^2 e^{-\frac{\arctan \left(\frac{-1+16 \gamma
 \Lambda +32 \gamma ^2 \Lambda \hat R(\bar t_0)}{\sqrt{-1+32 \gamma \Lambda }}\right)}{3 \sqrt{-1+32 \gamma \Lambda }}} \sqrt{4 \Lambda +\hat R(\bar t_0) \left(-1+16
\gamma \Lambda +16 \gamma ^2 \Lambda \hat R(\bar t_0)\right)}}{ (1+2 \gamma \hat R(\bar t_0))}$ with $\bar a_0$ as the present value of the scale factor, we get the first integral in the following form
\begin{multline}
\bar H(\bar t)^2+ \Lambda -\frac{\hat R(\bar t) (2+\gamma \hat R(\bar t))}{6 (1+2 \gamma \hat R(\bar t))^2}+\\
k\frac{e^{-\frac{\arctan \left(\frac{-1+16 \gamma
 \Lambda +32 \gamma ^2 \Lambda \hat R(\bar t)}{\sqrt{-1+32 \gamma \Lambda }}\right)}{3 \sqrt{-1+32 \gamma \Lambda }}} \sqrt{4 \Lambda +\hat R(\bar t) \left(-1+16
\gamma \Lambda +16 \gamma ^2 \Lambda \hat R(\bar t)\right)}}{C_0 (1+2 \gamma \hat R(\bar t))}=0.\label{firstintegral1}
\end{multline}
In consequence, the potential $V(\hat R)$ is given by
\begin{equation}
V(\hat R)=\frac{a^2}{2}\left(\Lambda -\frac{\hat R(\bar t) (2+\gamma \hat R(\bar t))}{6 (1+2 \gamma \hat R(\bar t))^2}\right)
\end{equation}
Because we know the form of $V(\hat R)$ and $\bar a(\hat R)$ we can get the potential $V(\bar a)$ in a numerical way. $V(\bar a)$ potential is demonstrated in Fig.~\ref{fig:10}.

Equations (\ref{dynsys1})--(\ref{dynsys2}) have the following the first integral given by
\begin{equation}
H(t)^2-\frac{(1+2 \gamma \hat R(t))^2 \left(-3 \Lambda +\hat R(t)-\frac{k (-4 \Lambda +\hat R(t))^{2/3}}{C_0}+\frac{\gamma (12 \Lambda
-3 \hat R(t)) \hat R(t)}{2 (1+2 \gamma \hat R(t))}\right)}{(1+2 \gamma \hat R(t)-3 \gamma (-4 \Lambda +\hat R(t)))^2} = 0,\label{firstintegral2}
\end{equation}
where $C_0=a_0^2 (-4 \Lambda +\hat R(t_0))^{2/3}$. Here, $a_0$ is the present value of the scale factor.

\section{Conclusions}

From detailed analysis of cosmological dynamics in the Palatini formulation we derive the following conclusions

\begin{enumerate}

\item If we consider the cosmic evolution in the Einstein frame we obtain inflation as an endogenous effect from dynamical formulation in the Palatini formalism \cite{Szydlowski:2017tbc}.

\item If we consider the cosmic evolution in the Jordan frame we obtain an exact and covariant formula for variability of gravitational constant $G_\text{eff}$ parametrized by the Ricci scalar.

\item Given two representations of our model in the Einstein and Jordan frames, we found that its dynamics is simpler in the Einstein frame as being free from some obstacles related with an appearance of bad singularities. It is an argument for the choice of the Einstein frame as physical.

\item In our model considered in the Einstein frame, we have both the inflation as well as the acceleration \cite{Szydlowski:2017tbc}. While the inflation in the model is obtained as an inherited dynamical effect, the acceleration is driven by the cosmological constant term.

\item In the model under consideration, we include effects of matter. This enable us to study the fragility of the inflation with respect to small changes of energy density of matter \cite{Szydlowski:2017tbc}.

\item In the obtained evolutional scenario of the evolution of the Universe we can unify: the singularity of the finite scale factor (generalized sudden singularity), the inflation with the sufficient number of e-folds and the phase of the acceleration of the current Universe \cite{Szydlowski:2017tbc}.

\item In the context of the Starobinsky model in the Palatini formalism we found a new type of double singularities beyond the well-known classification of isolated singularities.

\item The phase portrait for the Starobinsky model in the Palatini formalism with a positive value of $\gamma$ is equivalent to the phase portrait of the $\Lambda$CDM model. There is only a quantitative difference related with the presence of the non-isolated freeze singularity.

\item For the Starobinsky--Palatini model in the Einstein frame for the positive parameter $\gamma$, a sewn freeze singularity is replaced by a generalized sudden singularity. In consequence this model is not equivalent to the phase portrait of the $\Lambda$CDM model.

\end{enumerate}

\section*{Acknowledgements}
We are very grateful of A. Borowiec and A. Krawiec for stimulating discussion and remarks.

\providecommand{\href}[2]{#2}\begingroup\raggedright\endgroup


\begin{thebibliography}{10}

\bibitem{Sotiriou:2008rp}
T.~P. Sotiriou and V.~Faraoni, {\it {f(R) Theories of Gravity}}, Rev.
  Mod. Phys. {\bf 82} (2010) 451--497,
  \href{http://arxiv.org/abs/0805.1726}{{\tt arXiv:0805.1726}}.

\bibitem{Mukhanov:1981xt}
V.~F. Mukhanov and G.~V. Chibisov, {\it {Quantum Fluctuations and a Nonsingular
  Universe}},  JETP Lett. {\bf 33} (1981) 532--535. [Pisma Zh. Eksp.
  Teor. Fiz.33,549(1981)].

\bibitem{Starobinsky:1983zz}
A.~A. Starobinsky, {\it {The Perturbation Spectrum Evolving from a Nonsingular
  Initially De-Sitter Cosmology and the Microwave Background Anisotropy}},
  Sov. Astron. Lett. {\bf 9} (1983) 302.

\bibitem{Ade:2015rim}
{\bf Planck} Collaboration, P.~A.~R. Ade et~al., {\it {Planck 2015 results.
  XIV. Dark energy and modified gravity}}, Astron. Astrophys. {\bf 594}
  (2016) A14, \href{http://arxiv.org/abs/1502.01590}{{\tt arXiv:1502.01590}}.

\bibitem{Cheng:2013iya}
C.~Cheng, Q.-G. Huang, and Y.-Z. Ma, {\it {Constraints on single-field
  inflation with WMAP, SPT and ACT data — a last-minute stand before
  Planck}}, JCAP {\bf 1307} (2013) 018,
  \href{http://arxiv.org/abs/1303.4497}{{\tt arXiv:1303.4497}}.

\bibitem{Huang:2013hsb}
Q.-G. Huang, {\it {A polynomial f(R) inflation model}}, JCAP {\bf 1402}
  (2014) 035, \href{http://arxiv.org/abs/1309.3514}{{\tt arXiv:1309.3514}}.

\bibitem{Kofman:1985aw}
L.~A. Kofman, A.~D. Linde, and A.~A. Starobinsky, {\it {Inflationary Universe
  Generated by the Combined Action of a Scalar Field and Gravitational Vacuum
  Polarization}}, Phys. Lett. {\bf B157} (1985) 361--367.
  
\bibitem{Ketov:2010qz}
S.~V. Ketov and A.~A. Starobinsky, {\it {Embedding ($R+R^2$)-Inflation into
  Supergravity}}, Phys. Rev. {\bf D83} (2011) 063512,
  \href{http://arxiv.org/abs/1011.0240}{{\tt arXiv:1011.0240}}.

\bibitem{Appleby:2009uf}
S.~A. Appleby, R.~A. Battye, and A.~A. Starobinsky, {\it {Curing singularities
  in cosmological evolution of F(R) gravity}}, JCAP {\bf 1006} (2010)
  005, \href{http://arxiv.org/abs/0909.1737}{{\tt arXiv:0909.1737}}.

\bibitem{Capozziello:2009hc}
S.~Capozziello, M.~De~Laurentis, S.~Nojiri, and S.~D. Odintsov, {\it
  {Classifying and avoiding singularities in the alternative gravity dark
  energy models}}, Phys. Rev. {\bf D79} (2009) 124007,
  \href{http://arxiv.org/abs/0903.2753}{{\tt arXiv:0903.2753}}.

\bibitem{Alho:2016gzi}
A.~Alho, S.~Carloni, and C.~Uggla, {\it {On dynamical systems approaches and
  methods in $f(R)$ cosmology}}, JCAP {\bf 1608} (2016), no.~08 064,
  \href{http://arxiv.org/abs/1607.05715}{{\tt arXiv:1607.05715}}.

\bibitem{Capozziello:2015wsa}
S.~Capozziello, M.~F. De~Laurentis, L.~Fatibene, M.~Ferraris, and S.~Garruto,
  {\it {Extended Cosmologies}}, SIGMA {\bf 12} (2016) 006,
  \href{http://arxiv.org/abs/1509.08008}{{\tt arXiv:1509.08008}}.

\bibitem{Capozziello:2010sc}
S.~Capozziello, P.~Martin-Moruno, and C.~Rubano, {\it {Physical non-equivalence
  of the Jordan and Einstein frames}}, Phys. Lett. {\bf B689} (2010)
  117--121, \href{http://arxiv.org/abs/1003.5394}{{\tt arXiv:1003.5394}}.

\bibitem{Carroll:2004de}
S.~M. Carroll, A.~De~Felice, V.~Duvvuri, D.~A. Easson, M.~Trodden, and M.~S.
  Turner, {\it {The Cosmology of generalized modified gravity models}}, Phys. Rev. {\bf D71} (2005) 063513,
  \href{http://arxiv.org/abs/astro-ph/0410031}{{\tt astro-ph/0410031}}.

\bibitem{Borowiec:2011wd}
A.~Borowiec, M.~Kamionka, A.~Kurek, and M.~Szydlowski, {\it {Cosmic
  acceleration from modified gravity with Palatini formalism}},  JCAP
  {\bf 1202} (2012) 027, \href{http://arxiv.org/abs/1109.3420}{{\tt
  arXiv:1109.3420}}.

\bibitem{Olmo:2011uz}
G.~J.~Olmo, {\it {Palatini Approach to Modified Gravity: f(R) Theories and Beyond}}, Int. J. Mod. Phys.
  {\bf D20} (2011) 413--462, \href{http://arxiv.org/abs/1101.3864}{{\tt
  arXiv:1101.3864}}.
  
\bibitem{Olmo:2005hc}
G.~J.~Olmo, {\it {Post-Newtonian constraints on f(R) cosmologies in metric and Palatini formalism}}, Phys. Rev.
  {\bf D72} (2005) 083505, \href{http://arxiv.org/abs/gr-qc/0505135}{{\tt
  gr-qc/0505135}}.
  
\bibitem{Olmo:2005zr}
G.~J.~Olmo, {\it {The gravity Lagrangian according to solar system experiments}}, Phys. Rev. Lett.
  {\bf 95} (2005) 261102, \href{http://arxiv.org/abs/gr-qc/0505101}{{\tt
  gr-qc/0505101}}.
  
 \bibitem{Barragan:2010qb}
C.~Barragan and G.~J.~Olmo, {\it {Isotropic and Anisotropic Bouncing Cosmologies in Palatini Gravity}}, Phys. Rev.
  {\bf D82} (2010) 084015, \href{http://arxiv.org/abs/1005.4136}{{\tt
  arXiv:1005.4136}}.
  
\bibitem{Barragan:2009sq}
C.~Barragan, G.~J.~Olmo and H.~Sanchis-Alepuz {\it {Bouncing cosmologies in Palatini $f(R)$ gravity}}, Phys. Rev.
  {\bf D80} (2009) 024016, \href{http://arxiv.org/abs/0907.0318}{{\tt
  arXiv:0907.0318}}.
  
\bibitem{Bejarano:2017fgz}
C.~Bejarano, G.~J.~Olmo and D.~Rubiera-Garcia, {\it {What is a singular black hole beyond General Relativity?}}, Phys. Rev.
  {\bf D95} (2017) 064043, \href{http://arxiv.org/abs/1702.01292}{{\tt
  arXiv:1702.01292}}.
  
\bibitem{Bambi:2015zch}
C.~Bambi, A.~Cardenas-Avendano, G.~J.~Olmo and D.~Rubiera-Garcia, {\it {Wormholes and nonsingular spacetimes in Palatini $f(R)$ gravity}}, Phys. Rev.
  {\bf D93} (2016) 064016, \href{http://arxiv.org/abs/1511.03755}{{\tt
  arXiv:1511.03755}}.
  
\bibitem{Olmo:2015axa}
G.~J.~Olmo and D.~Rubiera-Garcia, {\it {Nonsingular Black Holes in $f(R)$ Theories}}, Universe
  {\bf 1} (2015) 173--185, \href{http://arxiv.org/abs/1509.02430}{{\tt
  arXiv:1509.02430}}.
  
\bibitem{Olmo:2011np}
G.~J.~Olmo and D.~Rubiera-Garcia, {\it {Nonsingular black holes in quadratic Palatini gravity}},  Eur. Phys. J.
  {\bf C72} (2012) 2098, \href{http://arxiv.org/abs/1112.0475}{{\tt
  arXiv:1112.0475}}.
  
\bibitem{Olmo:2011ja}
G.~J.~Olmo and D.~Rubiera-Garcia, {\it {Palatini $f(R)$ Black Holes in Nonlinear Electrodynamics}}, Phys. Rev.
  {\bf D84} (2011) 124059, \href{http://arxiv.org/abs/1110.0850}{{\tt
  arXiv:1110.0850}}.
  
\bibitem{Flanagan:2003rb}
E.~E.~Flanagan, {\it {Palatini form of $1/R$ gravity}}, Phys. Rev. Lett.
  {\bf 92} (2004) 071101, \href{http://arxiv.org/abs/astro-ph/0308111}{{\tt
  astro-ph/0308111}}.
  
\bibitem{Flanagan:2004bz}
E.~E.~Flanagan, {\it {The conformal frame freedom in theories of gravitation}}, Class. Quant. Grav.
  {\bf 21} (2004) 3817, \href{http://arxiv.org/abs/gr-qc/0403063}{{\tt
  gr-qc/0403063}}.
  
\bibitem{Pannia:2016qbj}
T.~Pannia et~al. {\it {Structure of Compact Stars in R-squared Palatini Gravity}}, Gen. Rel. Grav.
  {\bf 49} (2017) 25, \href{http://arxiv.org/abs/1607.03508}{{\tt
  arXiv:1607.03508}}. 
  
\bibitem{Koivisto:2005yk}
T.~Koivisto, {\it Covariant conservation of energy momentum in modified gravities}, Class. Quant. Grav. {\bf 23} (2006) 4289--4296, \href{http://arxiv.org/abs/gr-qc/0505128}{{\tt gr-qc/0505128}}.

\bibitem{DeFelice:2010aj}
A.~De~Felice and S.~Tsujikawa, {\it f(R) theories}, Living Rev. Rel. {\bf 13} (2010) 3, \href{https://arxiv.org/abs/1002.4928}{{\tt arXiv:1002.4928}}.
  
\bibitem{Olmo:2006eh}
G.~J.~Olmo, {\it 	
Limit to general relativity in f(R) theories of gravity}, Phys. Rev. {\bf D75} (2007) 023511, \href{http://arxiv.org/abs/gr-qc/0612047}{{\tt gr-qc/0612047}}.

\bibitem{Faraoni:2006hx}
V.~Faraoni, {\it Solar System experiments do not yet veto modified gravity models}, Phys. Rev. {\bf D74} (2006) 023529, \href{http://arxiv.org/abs/gr-qc/0607016}{{\tt gr-qc/0607016}}.

\bibitem{Szydlowski:2017uuy}
M.~Szydlowski, A.~Stachowski, A.~Borowiec, {\it 	
Emergence of running dark energy from polynomial f(R) theory in Palatini formalism},
Eur. Phys. J. {\bf C77} (2017) 603,
  \href{https://arxiv.org/abs/1707.01948}{{\tt arXiv:1707.01948}}.
  
\bibitem{Stachowski:2016dfi}
A.~Stachowski, M.~Szydlowski, {\it Dynamical system approach to running $\Lambda$ cosmological models},
Eur. Phys. J. {\bf C76} (2016) 606,
  \href{https://arxiv.org/abs/1601.05668}{{\tt arXiv:1601.05668}}.
  
\bibitem{Perko:2001de}
L.~Perko, \emph{Differential Equations and Dynamical Systems}, 3rd edn.
  (Springer, New York, 2001).
  
\bibitem{Szydlowski:2015fcq}
M.~Szydlowski, A.~Stachowski, A.~Borowiec, A.~Wojnar, {\it 	
Do sewn up singularities falsify the Palatini cosmology?},
Eur. Phys. J. {\bf C76} (2016) 567,
  \href{https://arxiv.org/abs/1512.04580}{{\tt arXiv:1512.04580}}.

\bibitem{Yurov:2017xjx}
A.~V. Yurov, A.~V. Astashenok, V.~A. Yurov,
{\it The cosmological models with jump discontinuities},
  \href{https://arxiv.org/abs/1710.05796}{{\tt arXiv:1710.05796}}.

\bibitem{Allemandi:2004wn}
G.~Allemandi, A.~Borowiec, M.~Francaviglia, {\it 	
Accelerated cosmological models in Ricci squared gravity},
Phys.Rev. {\bf D70} (2004) 103503,
\href{https://arxiv.org/abs/hep-th/0407090}{{\tt arXiv:hep-th/0407090}}.

	\bibitem{Nojiri:2005sx}
S.~Nojiri, S.~D.~Odintsov, and S.~Tsujikawa, {\it {Properties of singularities in (phantom) dark energy universe}},  Phys. Rev. {\bf D71} (2005) 063004, \href{http://arxiv.org/abs/hep-th/0501025}{{\tt hep-th/0501025}}.
	
	\bibitem{Barrow:2004xh}
J.~D.~Barrow, {\it Sudden future singularities},  Class. Quant. Grav. {\bf 21} (2004) L79-L82, \href{http://arxiv.org/abs/gr-qc/0403084}{{\tt gr-qc/0403084}}.
	
	\bibitem{BouhmadiLopez:2006fu}
M.~Bouhmadi-Lopez, P.~F.~Gonzalez-Diaz, and P.~Martin-Moruno, {\it {Worse than a big rip?}}, Phys. Lett. {\bf B659} (2008) 1--5,
\href{http://arxiv.org/abs/gr-qc/0612135}{{\tt gr-qc/0612135}}.
	
	\bibitem{Krolak:1986}
A.~Kr{\'o}lak, {\it {Towards the proof of the cosmic censorship hypothesis}}, Class. Quant. Grav. {\bf 3} (1986) 267-280.

	\bibitem{Sbierski:2015}
J.~Sbierski, {\it {The $C^0$-inextendibility of the Schwarzschild spacetime and the spacelike diameter in the Lorentzian Geometry}}, \href{http://arxiv.org/abs/1507.00601}{{\tt arXiv:1507.00601}}.

	\bibitem{Galloway:2016bej}
G.~J.~Galloway, E.~Ling {\it {Some remarks on the $C^0$-(in)extendibility of spacetimes}}, Annales Henri Poincare {\bf 18} (2017) 3427--3447, \href{http://arxiv.org/abs/1610.03008}{{\tt arXiv:1610.03008}}.

	\bibitem{Galloway:2017qkr}
G.~J.~Galloway, E.~Ling, J.~Sbierski, {\it {Timelike completeness as an obstruction to $C^0$-extensions}}, \href{http://arxiv.org/abs/1704.00353}{{\tt arXiv:1704.00353}}.

	\bibitem{Ling:2017uxe}
E.~Ling, {\it {Milne-like spacetimes and their role in cosmology}}, \href{http://arxiv.org/abs/1706.01408}{{\tt arXiv:1706.01408}}.

	\bibitem{Hrycyna:2008gk}
O.~Hrycyna, and M.~Szydlowski, {\it {Non-minimally coupled scalar field cosmology on the phase plane}}, {\em JCAP} {\bf 0904} (2009) 026, \href{http://arxiv.org/abs/0812.5096}{{\tt arXiv:0812.5096}}.
	
	\bibitem{Ellis:2015bag}
G.~F.~R.~Ellis, E.~Platts, D.~Sloan, and A.~Weltman, {\it {Current observations with a decaying cosmological constant allow for chaotic cyclic cosmology}}, JCAP {\bf 1604} (2016) 026, \href{http://arxiv.org/abs/1511.03076}{{\tt arXiv:1511.03076}}.

\bibitem{Dabrowski:2008kx}
M.~P.~Dabrowski, J.~Garecki, D.~B. Blaschke, {\it Conformal transformations and conformal invariance in gravitation}, Annalen
  Phys. {\bf 18} (2009) 13--32, \href{http://arxiv.org/abs/0806.2683}{{\tt
  arXiv:0806.2683}}.

\bibitem{Szydlowski:2017tbc}
M.~Szydlowski, A.~Stachowski, {\it 	
Simple cosmological model with inflation and late times acceleration}, \href{https://arxiv.org/abs/1708.04823}{{\tt arXiv:1708.04823}}.

\end{thebibliography}
\end{document}